\documentclass[a4paper,11pt]{article}
\pdfoutput=1 

\usepackage{jcappub} 

\usepackage[T1]{fontenc} 

\title{\boldmath Imprints of a galactic environment on TDEs: Suppressed peaks, shallower decays and late-time rebrightening episodes}

\author{Cedric Oña,}
\author{Sean Fortuna}
\author{and Ian Vega}

\affiliation{National Institute of Physics, University of the Philippines Diliman \\ Diliman, Quezon City 1101, Philippines}

\emailAdd{caona@nip.upd.edu.ph}
\emailAdd{sfortuna@nip.upd.edu.ph}
\emailAdd{ivega@nip.upd.edu.ph}

\abstract{Classic models of tidal disruption events (TDEs), employing a purely Keplerian description of stellar debris dynamics, have proven remarkably successful in describing the observed early-time emission of these transients. By construction, however, this picture treats the disruption as an isolated star-black hole encounter, leaving the gravitational influence of the surrounding galaxy absent from the dynamics. We relax this assumption by embedding the debris within the gravitational field of a spherically symmetric mass distribution representing either the host galaxy or a dark matter halo, while deliberately retaining a minimal Newtonian framework so that the environmental contribution remains transparent and straightforward to interpret. Within this extended model, we find that the host potential
can imprint clear signatures on the light curve that lie outside the predictive scope of traditional models, such as a suppressed initial accretion peak, phases of shallow decay and rebrightening episodes. For TDEs in ordinary galaxies, however, the latter signatures turns on far too late to be captured by current observations, explaining why simple Keplerian models have proven so effective despite neglecting the galactic environment. Only at much longer times does the broader galactic structure begin to reshape the fallback dynamics, gradually steering the system away from the canonical $t^{-5/3}$ decay. This provides an exciting opportunity for next-generation time-domain surveys to capture the long-term evolution of TDEs, enabling searches for delayed signals that encode information about the surrounding galactic mass.}

\begin{document}
\maketitle
\flushbottom


\section{Introduction} \label{sec:intro}

Supermassive black holes (SMBHs) are typically found at the heart of galaxies \citep{a, b}. Most of them, however, remain inactive and elusive to observation because they are not actively feeding on surrounding material. But occasionally, a violent process known as a tidal disruption event (TDE) reveals their presence whenever a star is accidentally nudged within the tidal disruption radius, where the black hole's tidal forces overwhelm the star's self-gravity, tearing it apart completely \citep{c, d}, or even only partially \citep{e} as in shallower encounters. Estimates suggest that only one event occurs every $10^4$ to $10^5$ years per galaxy \citep{f, g, h}, but when they do happen, they offer a unique opportunity to probe the otherwise hidden population of dormant black holes.

The earliest attempts to make sense of TDEs go back to the classic works of Lacy \textit{et al.} \citep{i}, Rees \citep{d} and Phinney \citep{j}. These early models posited a simple scenario in which a star on a nearly parabolic orbit is entirely disrupted, leaving behind a cloud of debris that settles into a wide distribution of orbits. About half of these stellar debris remain gravitationally bound and gradually return to the black hole, producing bright flares that can temporarily outshine the entire host galaxy \citep{k}. The rate at which the debris falls back, known as the fallback rate, is governed by the energy distribution at pericenter and follows a characteristic scaling law of $\dot{M}_{\mathrm{fb}} \propto t^{-5/3}$. More sophisticated semi-analytic treatments \citep{l, m, n} and early hydrodynamic simulations \citep{o, p, q, r} later provided detail, elucidating how a $t^{-5/3}$ fallback scaling can arise from the debris energy distribution in idealized, complete disruptions, while also showing that the temporal behavior of the fallback rate depends sensitively on stellar structure and encounter parameters.

This canonical fallback rate is typically used as a proxy for the observable luminosity, assuming that viscous accretion operates on a timescale much shorter than the fallback process itself \citep{s, t, u}. Under this assumption, the light curve of a TDE is shaped predominantly by the rate at which mass falls back onto the black hole. This has made the $t^{-5/3}$ decay law a widely used template for interpreting real TDE candidates \citep[e.g.][]{v, w, x}. 

But as wide-field time-domain surveys have come online, we have suddenly been flooded with data, and the picture has gotten messier (see \cite{y, z, aa} for a review). This influx of data has revealed a surprising diversity in light curve behavior. In response, theoretical efforts have expanded to explore how various parameters such as stellar structure \citep{q}, stellar rotation \citep{ab, ac}, orbital eccentricity \citep{ad, ae}, and impact parameter \citep{af, ag}, among others can shape the dynamics and outcomes of tidal disruptions.

Most existing studies focus on the local physics of the tidal disruption itself, where the influence of the black hole is expected to dominate. This is well justified, since the disruption occurs deep inside the black hole’s sphere of influence. However, the stellar debris does not evolve in isolation. As it streams outward and returns over long timescales, it remains embedded in the extended galactic potential, which acts as a persistent background field. Including this potential is therefore a simple yet insightful extension of standard fallback models. It allows us to revisit assumptions that are often taken for granted. Rather than presuming that the galactic field is entirely irrelevant, we can ask whether it plays any role at all, either during the disruption itself or in shaping the subsequent fallback and light curve at late times.

In this study, we embed the TDE within a more realistic galactic potential using a purely Newtonian approach. We model the galactic environment around the accreting black hole using a spherically symmetric double (broken) power-law density profile \citep{ah}, which captures the essential structural characteristics of various galactic formations, including globular clusters, galactic bulges, and dark matter halos \citep{ai, aj}. Our approach still leans on the analytical backbone of Rees \citep{d} and Lodato \textit{et al.} \citep{q}, but we move past the strictly Keplerian picture they employed.

\begin{figure}
    \centering
    \begin{minipage}{0.5\linewidth}
        \centering
        \includegraphics[width=\linewidth]{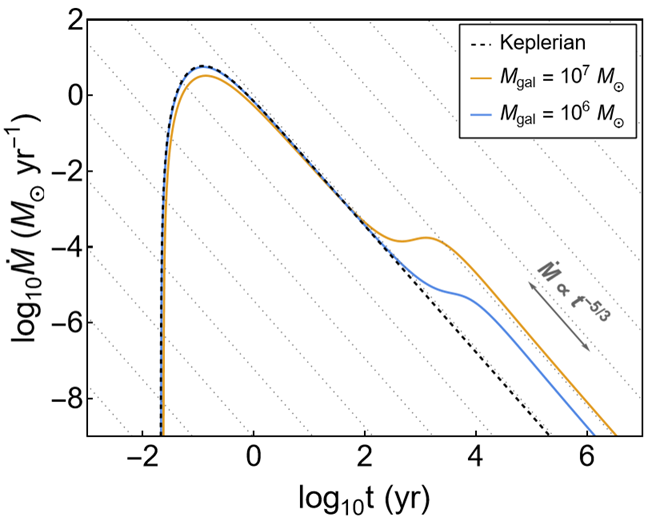}
    \end{minipage}\hfill
    \begin{minipage}{0.5\linewidth}
        \centering
        \includegraphics[width=\linewidth]{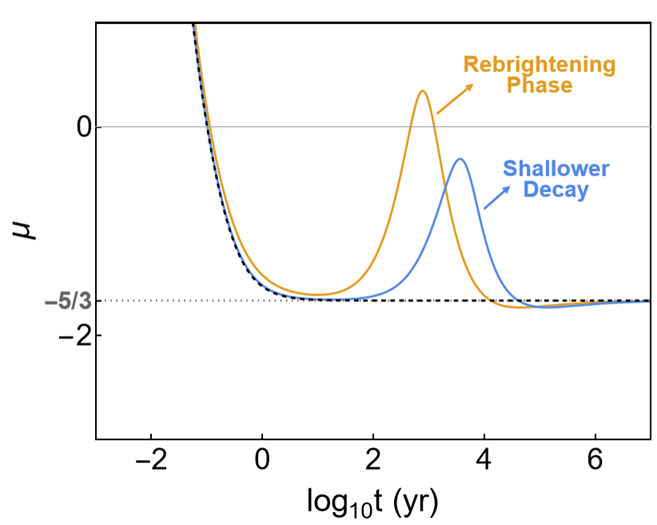}
    \end{minipage}
    \caption{Left: Time evolution of the mass accretion rate for two different IMBH-Hernquist systems. The black hole mass is set to $M_{\mathrm{bh}} = 10^4 \, M_{\odot}$, the scale length of the Hernquist environment to $b \approx 10^7 R_{\odot}$ and the encounter has a penetration factor $\delta = 1$. The star is solar-like and represented as a polytrope with index $n = 3$. The total galactic mass $M_{\mathrm{gal}}$ of the hosts are indicated in the plot legends. Right: Corresponding time evolution of the power-law index $\mu$ describing the decay slope. The region of shallower decay is characterized by a less negative slope ($-5/3 < \mu < 0$) whereas the rebrightening episode corresponds to a positive slope ($\mu > 0$).  For reference, the dotted lines in the background shows the canonical $t^{-5/3}$ decay slope expected in standard TDEs.}
    \label{fig:introfig}
\end{figure}

As a preview of our main findings, this extended yet deliberately simple setup, somewhat surprisingly, produces light curves with features absent in conventional treatments. Figure \ref{fig:introfig} illustrates how adding an external gravitational field drives the evolution away from the standard $t^{-5/3}$ decay, producing extended shallower-than-canonical decays and late-time rebrightening. 

Highly compact environments, as represented by the orange curve, can also affect the early-time energetics of the light curve by suppressing the initial peak accretion. More interestingly, over the 100-year timeframe, the light curve never touches the $t^{-5/3}$ decay. TDEs occurring in sufficiently compact environments can therefore exhibit persistently shallower decays throughout the observable window, suggesting a possible environmental origin for a subset of observed TDEs with shallow decay profiles. 

In contrast, more diffuse environments, such as that represented by the blue curve, are dynamically irrelevant in the early-time evolution, and the light curve closely follows the standard Keplerian fallback evolution over the 100-year window. This highlights that the effective compactness of the host environment determines whether deviations from the standard fallback behavior are expected to arise within practical observational windows.

At very late times, additional features emerge, including broad shallow tails and secondary peaks in the light curve. These signatures, however, arise only on timescales beyond those probed by most current observations, and at accretion levels that may fall below the sensitivity limits of our current instruments. Existing datasets are therefore only marginally sensitive, or in most cases completely blind, to these late-time signals, given how short most survey campaigns tend to be. Nevertheless, their very presence demonstrates that the host galaxy’s mass distribution can imprint itself directly onto the TDE light curve, revealing a deeper level of coupling between local and global gravitational influences than previously assumed. We will explore these behaviors in full detail in the later sections.

By moving beyond the classic Keplerian models, we get a richer picture of how TDEs might unfold in realistic environments. This approach not only enhances theoretical predictions but also carries implications for how we observe and interpret these rare events. In the long run, this means we might be able to use TDEs not just to probe black holes, but also to map the environments around them.

This paper is structured as follows. Section \ref{sec:model} lays out the extended model for TDEs in combined black hole-galaxy potentials. In Section \ref{sec:lc}, we present the resulting light curves and explore how different parameters shape the fallback signal. Section \ref{sec:par} maps out a more nuanced region of the parameter space and identifies conditions under which the light curves develop non-standard features, such as shallower-than-expected decay and rebrightening phases. Finally, we summarize our findings and draw our conclusions in Section \ref{sec:conc}.


\section{Extended TDE Model} \label{sec:model}

This section presents our extended TDE model, built under the combined gravitational influence of the central black hole and the surrounding galactic mass. We begin by constructing the system's total gravitational potential, laying the groundwork for the equations that govern the disruption process. We then outline the methodology, from setting up initial conditions to solving the equations that govern debris fallback.

\subsection{Constructing an extended black hole system}\label{sec:ebs}

\begin{figure}
    \centering
    \includegraphics[width=\linewidth]{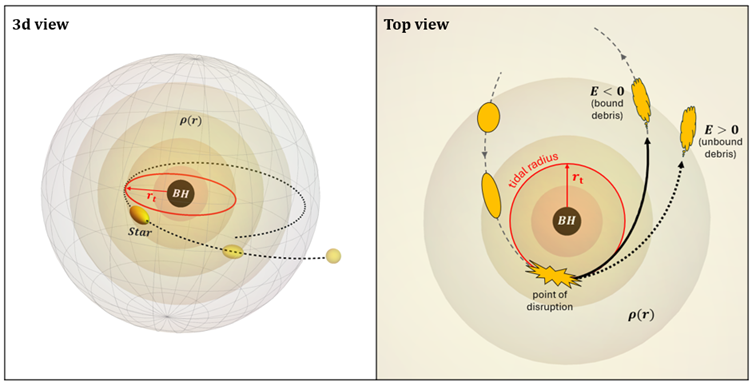}
    \caption{Schematic diagram of an extended black hole system. A central black hole (BH) is embedded within a spherically symmetric distribution of mass, represented by concentric shells whose density increases toward the center. The tidal radius $r_{\mathrm{t}}$ is shown in red, and the black dashed curve indicates the orbit of an approaching star. Upper panel: 3d view of the composite black hole-galaxy system. Lower panel: Top view showing the disruption process and the resulting debris stream. After the encounter, material with $E<0$ remains gravitationally bound to the black hole, while those with $E>0$ becomes unbound and escapes the system. This schematic is inspired by the illustrations presented in the works of Rees \citep{d} and Gezari \citep{ak}, with modifications to illustrate the extended host mass distribution.}
    \label{fig:ebs}
\end{figure}

We start with the gravitational potential of an isolated black hole. In the classical limit, this potential is governed by Newtonian gravity. For an isolated, non-rotating, spherically symmetric (Schwarszchild) black hole, this potential takes on a particularly simple form:
\begin{equation}
    \Phi_{\mathrm{bh}}(r) = -\frac{G M_{\mathrm{bh}}}{r}.
\end{equation}

However, real galaxies are far from empty. A black hole typically sits within a larger structure, whether it be a dense stellar cluster, a galactic bulge, or a dark matter halo. Therefore, we introduce an additional gravitational component sourced by this surrounding galactic mass. In this study, we only focus on spherically symmetric mass distributions, modeled by a double (broken) power-law density profile \citep{ah}. We choose this profile because it is flexible enough to capture both the steep inner regions and the shallower outer regions of realistic galactic environments, while still remaining mathematically simple to work with. This is expressed as
    \begin{equation} \label{eq:density}
        \rho(r)=\rho_0\left(\frac{r}{b}\right)^{-\gamma}\left[1+\left(\frac{r}{b}\right)^\alpha\right]^{(\gamma-\beta) / \alpha},
    \end{equation}
where $\gamma$ defines the inner slope of the galactic mass distribution, $\beta$ characterizes the outer slope, and $\alpha$ controls the sharpness of the transition between these two regions. The scale length $b$ sets the characteristic radius where the density profile transitions from the inner region to the outer envelope.

\begin{table}
 \caption{Two-power law density profiles commonly used in the literature and their corresponding structural parameters.}
 \label{tab:pot}
 \centering
 \begin{tabular}{c c c c c }
 \hline
  $(\alpha, \beta, \gamma)$ &  Name &  Potential, $\Phi(r)$ &  Density, $\rho(r)$ \\
  \hline \hline
  \\
  $(2,5,0)$ & Plummer \citep{al} & $-\dfrac{GM}{\sqrt{b^2+r^2}}$ & $\dfrac{3 M}{4\pi}\dfrac{b^2}{(b^2+r^2)^{5/2}}$\\
  \\
  $(1,4,1)$ & Hernquist \citep{am} & $-\dfrac{GM}{b+r}$ & $\dfrac{M}{2\pi}\dfrac{b}{r(b+r)^3}$ \\
  \\
  $(1,4,2)$ & Jaffe \citep{an} & $\dfrac{GM}{b}\ln\left|\dfrac{r}{b+r}\right|$ & $\dfrac{M}{4\pi}\dfrac{b}{r^2(b+r)^2}$\\
  \\
  \hline
 \end{tabular}
\end{table}

Figure \ref{fig:ebs} provides a visual representation of such an extended system. One can conceptually imagine this distribution as a series of concentric spherical shells that grow denser toward the center. The gravitational potential at radius $r$ due to a spherical mass distribution $\rho(r')$ is then determined by summing all the contributions from all the shells with $\mathrm{d} M\left(r^{\prime}\right)=4 \pi \rho\left(r^{\prime}\right) r'^2 \mathrm{d} r^{\prime}$ inside and outside of $r$. This is expressed mathematically as
\begin{equation}
    \Phi_{\mathrm{gal}}(r) = -4 \pi G \left[\frac{1}{r} \int_0^r  \rho\left(r^{\prime}\right) r^{\prime 2}\mathrm{d} r^{\prime}+ \int_r^{\infty}  \rho\left(r^{\prime}\right) r^{\prime}\mathrm{d} r^{\prime}\right]. \label{eq:shells}
\end{equation}

By evaluating the integrals in Equation \eqref{eq:shells}, we arrive at a closed-form expression for the gravitational potential of the surrounding mass. The final result, derived in full in Appendix \ref{sec:appA}, can be written using incomplete beta functions given by
\begin{equation}
    \begin{aligned}
        \Phi_{\mathrm{gal}}(r) =& -\frac{G M_{\mathrm{gal}}}{r}+ \frac{G M_{\mathrm{gal}}}{\mathcal{B}\left(\frac{\beta-3}{\alpha},\frac{3-\gamma}{\alpha}\right)} \left[\frac{1}{r}\mathcal{B}\left(\frac{\beta-3}{\alpha},\frac{3-\gamma}{\alpha},\mathcal{X}\right)\right.\\
        &\left.-\frac{1}{b}\mathcal{B}\left(\frac{\beta-2}{\alpha},\frac{2-\gamma}{\alpha},\mathcal{X}\right)\right], \label{eq:potmd}
    \end{aligned}
\end{equation}
where 
\begin{equation}
    \mathcal{X} = \frac{1}{1+\left(\frac{r}{b}\right)^{\alpha}}.
\end{equation}

In several cases, Equation \eqref{eq:potmd} reduces to a cleaner analytical expression. Table \ref{tab:pot} lists several of these special cases, which are commonly used in the literature for their renowned ability to approximate the characteristics of diverse galactic environments. For this study, we only focus on finite mass potentials, specifically those satisfying $\gamma < 3$ and $\beta > 3$, including the widely used models like Hernquist, Jaffe and the Plummer sphere. Among these, we often refer to the Hernquist potential as our representative case.

The total gravitational potential of the extended system is then given by the sum of the contributions of the two components, that is
\begin{equation}
    \Phi_{\mathrm{tot}}(r) = \Phi_{\mathrm{bh}}(r) + \Phi_{\mathrm{gal}}(r).
\end{equation}
Explicitly, we have
\begin{equation}
    \begin{aligned}
    \Phi_{\mathrm{tot}}(r) =& -\frac{G (M_{\mathrm{bh}}+M_{\mathrm{gal}})}{r}+ \frac{G M_{\mathrm{gal}}}{\mathcal{B}\left(\frac{\beta-3}{\alpha},\frac{3-\gamma}{\alpha}\right)} \left[\frac{1}{r}\mathcal{B}\left(\frac{\beta-3}{\alpha},\frac{3-\gamma}{\alpha},\mathcal{X}\right)\right.\\
    &\left.-\frac{1}{b}\mathcal{B}\left(\frac{\beta-2}{\alpha},\frac{2-\gamma}{\alpha},\mathcal{X}\right)\right].
\end{aligned}
\end{equation}

\subsection{TDE dynamics and basic equations}\label{sec:TDEdyn}

The process by which a star is captured and brought close enough to the black hole to be disrupted is typically studied under loss cone theory \citep{f, g, h}, which describes the range of orbital parameters that lead to a star being disrupted rather than simply orbiting the black hole. The present work does not address the early scattering processes that place stars onto disruptive orbits. Instead, we start at the precise moment the star is fully disrupted by tidal forces, following the earlier analysis of Rees \citep{d} and later refinements made by Lodato \textit{et al.} \citep{q}. Our main interest is the evolution of the debris, focusing on how the liberated gas moves under the combined pull of the black hole and the surrounding galactic mass.

To begin, we define the tidal field as the differential in the gravitational field across the stellar radius. We have
\begin{equation}
    F_{\mathrm{tid}}(r) = \frac{\mathrm{d}\Phi_{\mathrm{tot}}(r)}{\mathrm{d}r} - \frac{\mathrm{d}\Phi_{\mathrm{tot}}(r+R_{*})}{\mathrm{d}r}.
\end{equation}

Tidal disruption occurs when this tidal force exceeds the star's self-gravity. The disruption condition can thus be written as
\begin{equation}
    \frac{\mathrm{d}\Phi_{\mathrm{tot}}(r)}{\mathrm{d}r} - \frac{\mathrm{d}\Phi_{\mathrm{tot}}(r+R_{*})}{\mathrm{d}r} = \frac{G M_{*}}{R_{*}^2}. \label{eq:tr condition}
\end{equation}

Solving the above equation for $r$ yields the tidal radius $R_{\mathrm{t}}$, the critical distance at which the star becomes vulnerable to complete disruption. In some encounters, the star plunges far deeper than this critical radius. To quantify the depth of this encounter, we define the penetration factor $\delta$ as
\begin{equation}
    \delta = \frac{R_{\mathrm{t}}}{R_{\mathrm{p}}},
\end{equation}
where $R_{\mathrm{p}}$ is the pericenter distance.

In this work, we adopt the standard assumption that the star approaches the black hole on a parabolic, zero-energy trajectory, a common situation in galactic centers where frequent stellar encounters scatter stars into nearly parabolic orbits before disruption \citep{ao}. This allows us to approximate the energy spread imparted to the debris by
\begin{align}
    \Delta E \approx \left(\frac{\mathrm{d}U_{\mathrm{tot}}}{\mathrm{d}r}\right)_{R_{\mathrm{p}}} R_{*},
\end{align}
where $U_{\mathrm{tot}}$ is the potential energy due to the extended black hole system. This energy spread gives rise to a spectrum of trajectories for the stellar debris, wherein a fraction of the stripped material ($E > 0$) is expelled from the system, while the remainder ($E < 0$) remains gravitationally bound to the black hole.

Each bound stellar debris is then tracked based on its specific energy and orbital period. Our model employs the frozen-in approximation, which assumes that at the tidal radius (or pericenter), the star’s structure is "frozen" and each fluid element instantly acquires its orbital energy, after which the debris evolves ballistically in the black hole’s gravity without further influence from the star’s self-gravity or pressure forces \citep{ap}. Here, however, we extend this picture by allowing the debris to respond not only to the black hole, but also to the gravitational pull of the surrounding host environment.

We estimate debris trajectories using the circular velocity, given by
\begin{equation}
    v_c(r) = \left(r\frac{\mathrm{d}\Phi_{\mathrm{tot}}}{\mathrm{d}r}\right)^{1/2}. \label{eq:vc}
\end{equation}
From this, the orbital period is simply
\begin{equation}
    T(r) = \frac{2\pi r}{v_c(r)}, \label{eq:period}
\end{equation}
and the specific orbital energy is given by
\begin{align}
    E(r) = \frac{1}{2}v_c^2(r) + \Phi_{\mathrm{tot}}(r). \label{eq:energy}
\end{align}

Assuming the fallback material is efficiently accreted, the distribution of return times can be used to estimate the mass accretion rate $\dot{M}_{\mathrm{acc}}$. Using standard energy-time mapping, this feeding rate can be expressed as
\begin{align}
    \frac{\mathrm{d}M}{\mathrm{d}T} = \frac{\mathrm{d}M}{\mathrm{d}E}\frac{\mathrm{d}E}{\mathrm{d}T}, \label{eq:mar}
\end{align}
which in turn determines the radiative output via \citep{aq}
\begin{align}
    \dot{M}_{\mathrm{acc}} = \frac{L_{\mathrm{acc}}}{\eta c^2},
\end{align}
where $\eta$ is the radiative efficiency of the accretion process. 

This accretion luminosity, however, cannot grow arbitrarily large. Above a certain threshold, called the Eddington luminosity, radiation pressure balances gravity \citep{ar}, halting further inflow of matter. For thin-disk accretion with 
$\eta \approx 0.1$, the Eddington limit is approximately equal to
\begin{align}
    \dot{M}_{\mathrm{edd}} = \frac{L_{\mathrm{edd}}}{\eta c^2} \approx 10^{-8} \left(\frac{M_{\mathrm{h}}}{M_{\odot}}\right) \frac{M_{\odot}}{\mathrm{yr}}. \label{eq:edd}
\end{align}

Following the approach of Lodato \textit{et al.} \citep{q}, we model the disrupted star as a polytropic sphere. This helps us describe the mass distribution over specific energies, $\mathrm{d}M/\mathrm{d}E$, using the star's internal density structure:
\begin{equation}
     \frac{\mathrm{d}M}{\mathrm{d}E} = \frac{\mathrm{d}M}{\mathrm{d}\Delta r}\frac{R_*}{\Delta E}. \label{eq:spread}
\end{equation}

One can visualize the disrupted star as being "salami sliced" \citep{as}, where each slice is $\Delta r$ away from the center. Each thin cylindrical slab obtains the same specific energy at the pericenter. The mass distribution along this radial distance $\mathrm{d}M/\mathrm{d}\Delta r$ can be computed via
\begin{equation}
    \frac{\mathrm{d}M}{\mathrm{d}\Delta r} = 2\pi \int_{\Delta r}^{R_{*}} \rho(r') r' dr', \label{eq:dist}
\end{equation}
where $\rho(r)$ is the spherically symmetric density of the star, obtained by solving the Lane-Emden equation.

For the purposes of this paper, we only focus on the case of a polytropic index $n=3$, which crudely represents solar-type main sequence stars \citep{aq}. The Lane-Emden solution for $n=3$ gives us a well-defined density profile that feeds directly into the calculation of $\mathrm{d}M/\mathrm{d}E$, and ultimately, the mass fallback rate $\mathrm{d}M/\mathrm{d}T$.

\subsection{Setup and exact solutions} \label{sec:setup}

Traditional Keplerian TDE models inherently revolve around six key parameters \citep{af}, which are the stellar mass $M_{*}$, the stellar radius $R_{*}$, the black hole mass $M_{\mathrm{bh}}$, the orbital eccentricity $e$, the polytropic index $n$, and the penetration factor $\delta$. However, when we introduce the influence of  the specific extended mass distribution, the situation becomes considerably more complex. This addition brings with it five additional parameters that must be taken into account. These include the total mass of the surrounding mass distribution $M_{\mathrm{gal}}$, the characteristic scale length $b$ of that distribution, and three structural parameters ($\alpha$, $\beta$, $\gamma$) that together characterize the density profile of the galactic mass.

To simplify the analysis and reduce the dimensionality of the problem, we held some of the parameters fixed in our simulations. First, we model the disrupted star as a Sun-like star, setting its radius and mass to $R_{*}=R_{\odot}$ and $M_{*}=M_{\odot}$, respectively. We also assume a polytropic index of $n=3$, and consider the approaching star to be on a parabolic orbit, eliminating explicit dependence on eccentricity.

For the galactic component, we adopt the Hernquist potential, with $(\alpha, \beta, \gamma)_{\mathrm{hern}} = (1,4,1)$, as our default model. This potential is a popular choice for describing stellar bulges due to its analytic tractability and empirical fit to observed galactic cores \citep{at, au}. To check that our results are not unduly influenced by this particular choice, we also consider two alternative finite mass potentials: the Jaffe and Plummer spheres. These models are characterized by the structural parameters $(\alpha, \beta, \gamma)_{\mathrm{jaf}} = (1,4,2)$ and $(\alpha, \beta, \gamma)_{\mathrm{plum}} = (2,5,0)$, respectively.

With these constraints in place, our analysis is effectively distilled to a 4-dimensional parameter space consisting of $M_{\mathrm{bh}}, M_{\mathrm{gal}}, b$ and $\delta$.

To make calculations simpler, we shift to a set of dimensionless quantities listed in Table \ref{tab:nd}. Physical variables are normalized using the Sun’s radius and mass as reference scales. Time is normalized via
\begin{equation}
    T_0 = 2\pi \left(\frac{R_{\mathrm{p}}^3}{G M_{\mathrm{bh}}}\right)^{1/2},
\end{equation}
and density by
\begin{equation} 
    \rho_0 = \frac{M_{*}}{R_{*}^3}. \label{eq:fidrho}
\end{equation}
From here on, all equations are written in this dimensionless variables.

By recasting Equations \eqref{eq:vc}, \eqref{eq:period}, and \eqref{eq:energy} using these variables, we derive expressions for the dimensionless orbital period and specific energy. Table \ref{tab:orb} summarizes these results for the three finite galactic potentials that we consider in our analysis (Hernquist, Jaffe, and Plummer), along with the purely Keplerian case of an isolated black hole for reference.

\begin{table}
 \caption{Dimensionless quantities used in our extended TDE model, adapted from Lodato \textit{et al.} \citep{q} with additional definitions included. Each quantity is normalized by a characteristic astrophysical scale, such as the solar mass, solar radius, or appropriate energy, time, and density scales.}
 \label{tab:nd}
 \centering
 \begin{tabular}{l c l}
 \hline
  Quantity & & Dimensionless form\\
  \hline \hline
  Black hole mass & &$m_{\mathrm{bh}}=M_{\mathrm{bh}}/M_{*}$ \\
  Galactic mass & &$m_{\mathrm{gal}}=M_{\mathrm{gal}}/M_{*}$ \\
  Tidal radius & & $r_{\mathrm{t}} = R_{\mathrm{t}}/R_{*}$ \\
  Pericenter distance & & $r_{\mathrm{p}} = R_{\mathrm{p}}/R_{*}$\\
  Scale length & & $s=b/R_{*}$ \\
  Radial distance from the black hole & & $x=r/R_{*}$ \\
  Stellar radial coordinate & & $\xi=\Delta r/R_{*}$ \\
  Energy & & $\epsilon = -E/\Delta E$\\
  Time & & $\tau = T/T_0$ \\
  Density & & $\hat{\rho} = \rho/\rho_0$ \\
  \hline
 \end{tabular}  
\end{table}

The analysis starts by calculating the tidal radius from Equation \eqref{eq:tr condition}. It provides the natural scale against which the pericenter distance $r_{\mathrm{p}}$ is measured, thereby defining the penetration factor of the encounter.

Once the point of disruption is determined, the next step is to establish the energy-period relation, which we denote as $\epsilon(\tau)$, and then compute the energy rate, $\mathrm{d}\epsilon/\mathrm{d}\tau$. In the purely Keplerian case, this relation is straightforward and can be written analytically. But with an extended galactic mass, things get more complicated, and we must turn to numerical methods. The energy rate is then calculated using
\begin{equation}
    \frac{\mathrm{d}\epsilon}{\mathrm{d}\tau} = \frac{\mathrm{d}\epsilon/\mathrm{d}x}{\mathrm{d}\tau/\mathrm{d}x}, \label{eq:dedt}
\end{equation}
where the derivatives are taken with respect to the radial coordinate $x$.

We define the range of $x$ based on a "salami-slice" decomposition of the star. The global coordinate $x$ tracks the distance of each slice from the black hole itself.

We also define an internal stellar coordinate $\xi = \Delta r/R_*$, where $\Delta r$ is the distance from the stellar center to a given salami slice. By convention, $\Delta r > 0$ is measured toward the black hole, such that the near-side surface of the star has the largest positive $\Delta r$, while $\Delta r = 0$ corresponds to the stellar center. 

The slice nearest the black hole is assigned $\epsilon = 1$ by construction, corresponding to the most tightly bound debris, which is most susceptible to tidal stripping and responsible for early fallback. The slice at the stellar center is characterized by $\epsilon = 0$, representing the least bound material that contributes to the later stages of accretion. The far-side layers of the star with $\epsilon < 0$ are considered unbound and are assumed to be expelled from the system without contributing to the accretion process.

We determine these energy bounds numerically using the specific energy expressions provided in Table \ref{tab:orb}. To prevent computational issues near the singular point at exactly $\epsilon = 0$, we impose a small lower cutoff at $\epsilon = 10^{-8}$. Once the upper and lower limits in $x$ are established, we discretize this range into finite intervals and evaluate $\mathrm{d}\epsilon/\mathrm{d}\tau$ across this grid.

\begin{table}
 \caption{Orbital period and specific energy of the disrupted stellar debris for different galactic environments, expressed in terms of dimensionless quantities. The table includes results for an isolated black hole (purely Keplerian) and three extended galactic potentials: Hernquist $+$ BH, Jaffe $+$ BH, and Plummer $+$ BH.}
 \label{tab:orb}
 \centering
 \scriptsize
 \begin{tabular}{ c c c }
 \hline
  System & Orbital period, $\tau(x)$ & Specific energy, $\epsilon(x)$ \\
  \hline \hline
  Isolated BH & $\left(\dfrac{x}{r_{\mathrm{p}}}\right)^{3/2}$ & $\dfrac{r_{\mathrm{p}}^2}{2x}$
  \\
  \\
  BH + Hernquist & $\sqrt{\dfrac{m_{\mathrm{bh}}(s+x)^2}{m_{\mathrm{bh}}(s+x)^2+m_{\mathrm{gal}}x^2}}\left(\dfrac{x}{r_{\mathrm{p}}}\right)^{3/2}$ & $\dfrac{r_{\mathrm{p}}^2}{2}\left(\dfrac{1}{x}+\dfrac{m_{\mathrm{gal}}(2s+x)}{m_{\mathrm{bh}}(s+x)^2}\right)$
  \\
  \\
  BH + Jaffe & $\sqrt{\dfrac{m_{\mathrm{bh}}(s+x)}{m_{\mathrm{bh}}(s+x)+m_{\mathrm{gal}}x}}\left(\dfrac{x}{r_{\mathrm{p}}}\right)^{3/2}$ & $\dfrac{r_{\mathrm{p}}^2}{2}\left(\dfrac{1}{x} - \dfrac{m_{\mathrm{gal}}}{m_{\mathrm{bh}}(s+x)} - \dfrac{2 m_{\mathrm{gal}}\ln\left|\frac{x}{s+x}\right|}{m_{\mathrm{bh}}s}\right)$
  \\
  \\
  BH + Plummer & $\sqrt{\dfrac{m_{\mathrm{bh}}(s^2+x^2)^2}{m_{\mathrm{bh}}(s^2+x^2)^2+m_{\mathrm{gal}}x^3\sqrt{s^2+x^2}}}\left(\dfrac{x}{r_{\mathrm{p}}}\right)^{3/2}$ & $\dfrac{r_{\mathrm{p}}^2}{2}\left(\dfrac{1}{x}+\dfrac{m_{\mathrm{gal}}(2s^2+x^2)}{m_{\mathrm{bh}}(s^2+x^2)^{3/2}}\right)$
  \\
  \hline
 \end{tabular} 
\end{table}

After that, we proceed by obtaining the mass distribution of specific energies, $\mathrm{d}m/\mathrm{d}\epsilon$. This requires solving the Lane-Emden equation for $n=3$, and then evaluating the dimensionless versions of Equations \eqref{eq:spread} and \eqref{eq:dist}:
\begin{equation}
    \frac{\mathrm{d}m}{\mathrm{d}\epsilon} =
    \frac{\mathrm{d}m}{\mathrm{d}\xi} = 2\pi \int_{
    \xi}^{1} \hat{\rho} (\xi') \xi' \mathrm{d}\xi'. \label{eq:distnd}
\end{equation}

Equation \eqref{eq:distnd} can be recast into a more convenient form as
\begin{align}
    \frac{\mathrm{d}m}{\mathrm{d}\epsilon} = \zeta_1 \frac{\mathrm{d}m}{\mathrm{d}\zeta},
\end{align}
where $\zeta_1$ denotes the dimensionless stellar radius obtained from the Lane–Emden equation. The quantity $\mathrm{d}m/\mathrm{d}\zeta$ is then determined by solving the differential equation
\begin{align}
    y'(\zeta) - \frac{1}{2}\frac{\zeta \theta^n(\zeta)}{\zeta_1\left(\frac{\mathrm{d}\theta}{\mathrm{d}\zeta}\right)_{\zeta_1}} = 0,
\end{align}
with $y(\zeta) \equiv \mathrm{d}m/\mathrm{d}\zeta$ and $\theta(\zeta)$ denotes the Lane–Emden density solution. This equation is solved numerically subject to the boundary condition $y(\zeta_1) = 0$.

With both $\mathrm{d}m/\mathrm{d}\epsilon$ and $\mathrm{d}\epsilon/\mathrm{d}\tau$ in hand, we finally get the mass accretion rate by simply taking their product,
\begin{equation}
    \frac{\mathrm{d}m}{\mathrm{d}\tau} = \frac{\mathrm{d}m}{\mathrm{d}\epsilon} \frac{\mathrm{d}\epsilon}{\mathrm{d}\tau}.
\end{equation}

Additionally, we monitor how the accretion rate evolves over time by tracking the instantaneous power-law index $\mu$ given by
\begin{equation}
    \mu = \frac{\mathrm{d}\ln \dot{m}}{\mathrm{d}\ln\tau}.
\end{equation}

The classical model predicts that at late times, the mass accretion rate follows a power-law decay, $\dot{m} \propto \tau^{-\mu}$, with $\mu \approx 5/3$ for full disruptions. Tracking the temporal evolution of $\mu$ provides a clear way to detect deviations from this standard behavior, allowing us to identify transitions between accretion phases, late-time departures from the expected decay, and potential rebrightening episodes.


\section{TDE Light Curves} \label{sec:lc}

In this section, we look at how the mass accretion rate and its power-law index evolve over time once the extended potential is taken into account. We pay special attention to four important parameters: the black hole mass $m_{\mathrm{bh}}$, the total mass of the surrounding galactic environment $m_{\mathrm{gal}}$, its scale radius $s$ and penetration factor $\delta$. Throughout this section, we only consider galactic environments modeled by the Hernquist potential.

\subsection{Across different black hole masses} \label{sec:bhmass}

\begin{figure*}
    \centering
    \begin{minipage}{0.5\linewidth}
        \centering
        \includegraphics[width=\linewidth]{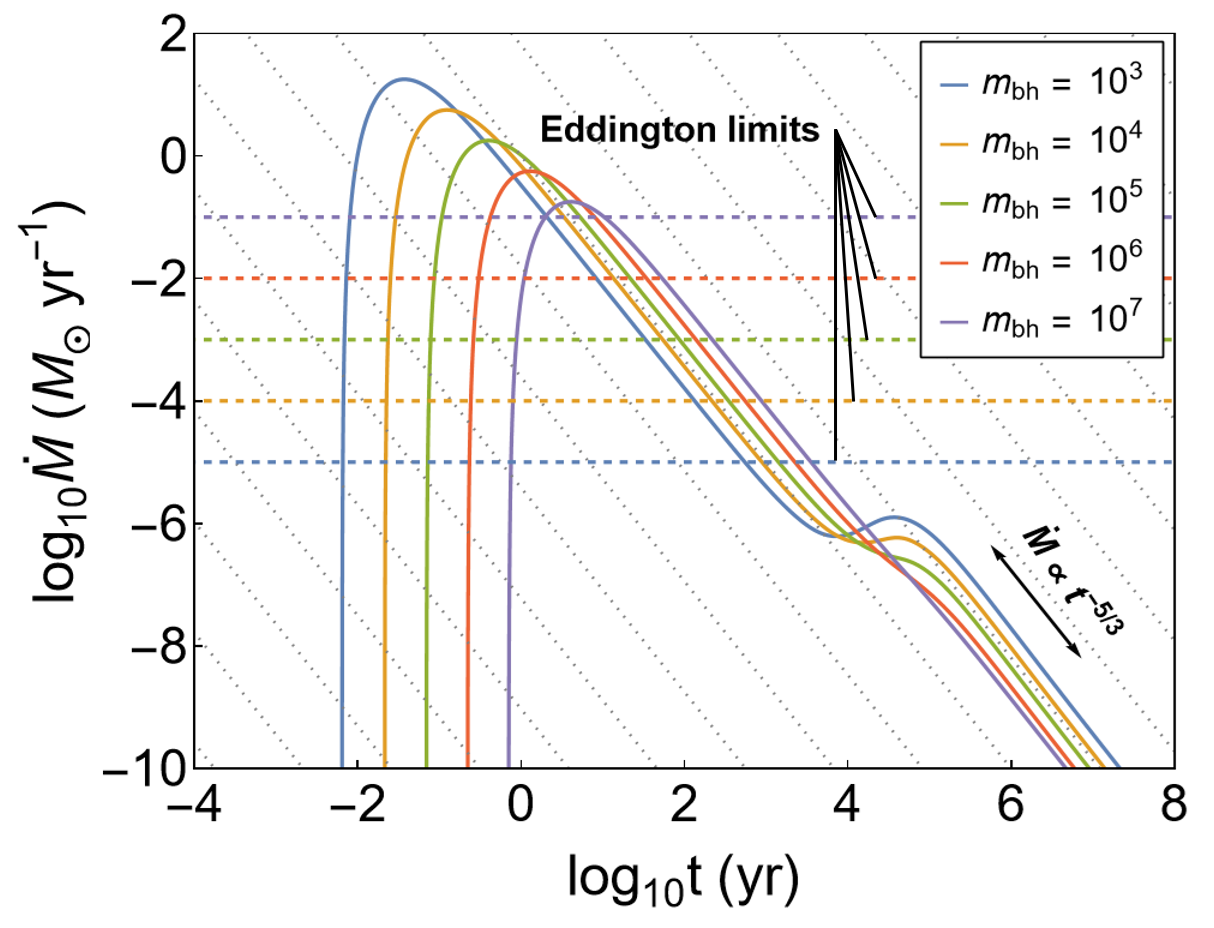}
    \end{minipage}\hfill
    \begin{minipage}{0.5\linewidth}
        \centering
        \includegraphics[width=\linewidth]{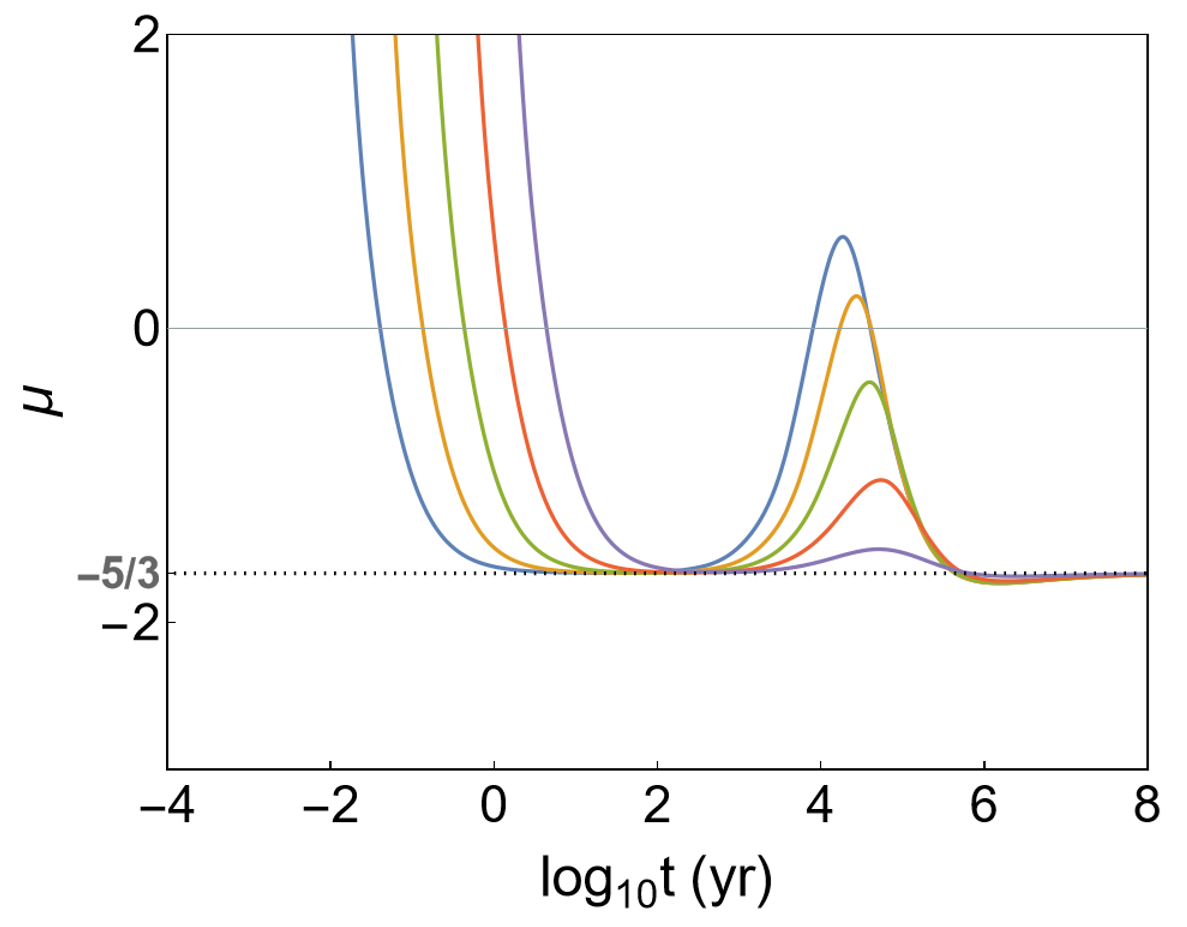}
    \end{minipage}
    \caption{Left: Time evolution of the mass accretion rate for a range of black hole masses. The black hole is embedded in a Hernquist environment with a fixed galactic mass of $10^7 M_{\odot}$ and scale radius of $10^8 R_{\odot}$. The penetration factor is set to $\delta = 1$. Dashed lines indicate the Eddington limit corresponding to each black hole mass, while the dotted background curve shows the canonical fallback rate scaling of $t^{-5/3}$ for reference. Right: Corresponding evolution of the instantaneous power-law index. The dotted line marks the standard $-5/3$ slope, and the solid horizontal line indicates a slope of $0$, distinguishing light curves with secondary peaks (rebrightening phases) from those with broad shallow tails (shallower-than-standard decay).}
    \label{fig:bhmass}
\end{figure*}

Figure \ref{fig:bhmass} illustrates how the mass accretion rate and the corresponding power-law index evolve over time for a range of black hole masses. We embed the black hole in a NSC-like environment, motivated by the expectation that more compact galactic settings exert stronger gravitational influence on the debris fallback dynamics. We adopt a scale radius of approximately 3 parsecs, which translates to a dimensionless value of $s \sim 10^8$. This choice is consistent with the observed half-light radii of NSCs, which typically span around $2$ to $5$ parsecs across a broad spectrum of galaxy types \citep{av, aw}. For a Hernquist profile, the half-light radius and scale radius are related by $r_h = 1.815 \, b$ \citep{ax}. Despite their small size, NSCs are known to be massive, with dynamical masses generally falling within the range of $10^6 - 10^8 \, M_\odot$ \citep{ay, az}. Thus, we take $m_{\mathrm{gal}} = 10^7$ as a representative value.

For this setup, we explore the fallback behavior for black holes with masses ranging from $10^3$ to $10^7 \, M_\odot$, spanning both IMBH and low-end SMBH regimes. Note that for black holes with mass $M_{\mathrm{bh}} \gtrsim 10^7 \, M_\odot$, tidal disruption of main-sequence stars occurs close enough to the event horizon that a Newtonian treatment of the tidal field is no longer valid \citep{ba}.

The left panel of Figure \ref{fig:bhmass} shows that the black hole mass primarily regulates the peak accretion rate and flare timing. Increasing $m_{\mathrm{bh}}$ reduces the peak rate by roughly an order of magnitude and delays the onset of the flare, consistent with established theoretical expectations \citep{bb, ak}. This agreement indicates that the inclusion of an extended galactic potential does not significantly modify the canonical early-time dependence of fallback on black hole mass, at least with our current setup and choice of environment.

At later times, however, secondary features emerge in the form of shallow tails and rebrightening phases. But such signatures arise only after the accretion rate has dropped by several orders of magnitude into the sub-Eddington regime, placing them at the edge, or entirely beyond the reach of current survey cadences and sensitivity limits. This naturally explains why most observed TDEs exhibit smooth, monotonic declines consistent with the expected $t^{-5/3}$ scaling.

The right panel reveals these deviations more clearly through the evolution of the instantaneous power-law index. While the curves initially converge toward the canonical value of $-5/3$, they later become shallower and may even turn positive, signaling rebrightening episodes, before eventually returning to the standard decay at very late times. In typical observational windows, only the early convergence toward the canonical slope is likely to be detected.

Furthermore, Figure \ref{fig:bhmass} shows that environmental effects are most pronounced for IMBH systems ($10^3$–$10^5 \, M_\odot$). This trend reflects the fact that, for lower-mass black holes, the gravitational influence of the surrounding galactic mass becomes comparatively more competitive with that of the black hole during the fallback evolution.

In practice, however, many of these signatures emerge on timescales far beyond those accessible to current observations, making their direct detection challenging. Detectability may therefore be limited to exceptionally compact environments, where the external potential induces stronger and earlier deviations from the canonical evolution. Although difficult to capture at present, such signals could in principle be recovered through long-baseline monitoring or cross-survey identification of late-time emission. Future time-domain facilities with improved sensitivity and extended coverage will therefore play a crucial role in testing these predictions.

\subsection{Influence of the host galaxy} \label{sec:host}

Here, we investigate how the properties of the host galaxy influence the fallback evolution by varying two key parameters that characterize the surrounding potential: the total galactic mass $m_{\mathrm{gal}}$ and its scale length $s$. Together, these parameters control the spatial distribution of mass around the black hole and thus set the effective compactness of the host environment.


\subsubsection{Across varying total galactic masses} \label{sec:galmass}

Previously, we showed that lower-mass black holes allow environmental effects to influence fallback dynamics more strongly. Motivated by this result, we set the black hole mass to $m_{\mathrm{bh}} = 10^4$, representative of IMBHs, for all simulations in this and the following subsections. We then compute light curves for a range of galactic masses while holding the scale length fixed at $s = 10^8$. The galactic component is chosen to resemble a nuclear star cluster, with masses spanning $m_{\mathrm{gal}} = 10^6 - 10^8$.

\begin{figure*}
    \centering
    \begin{minipage}{0.5\linewidth}
        \centering
        \includegraphics[width=\linewidth]{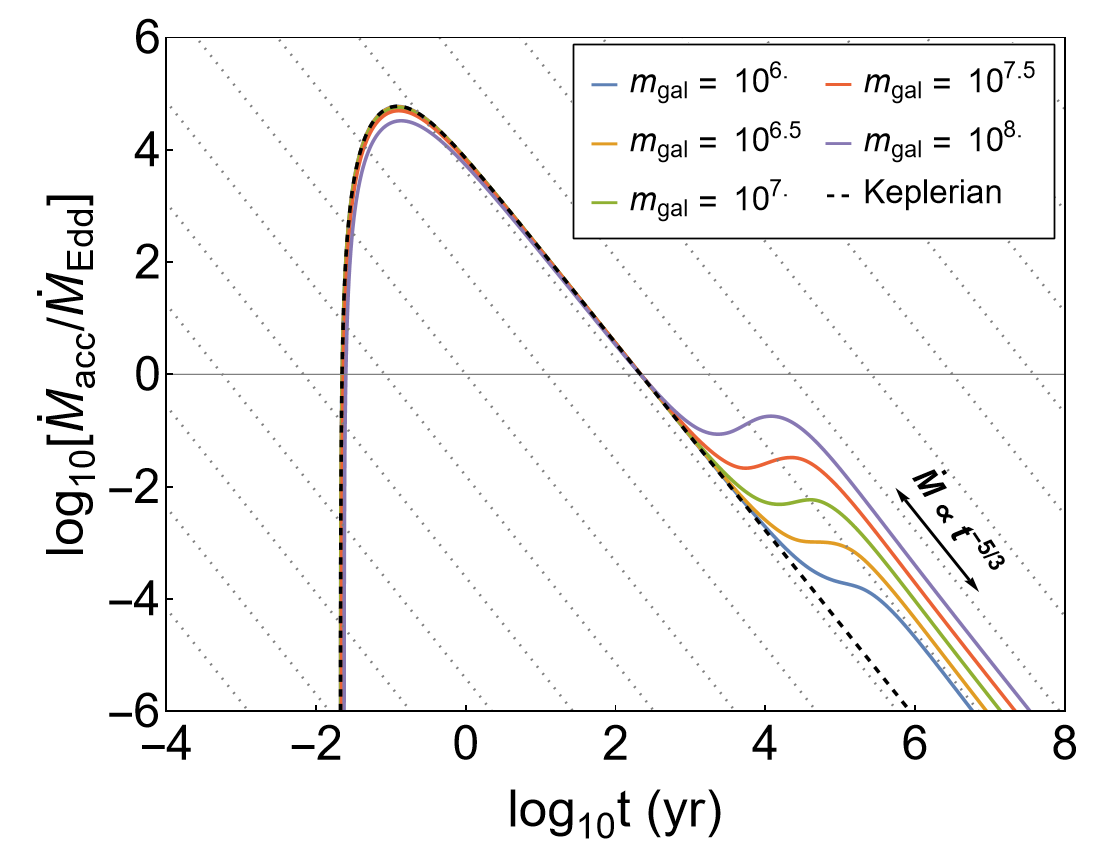}
    \end{minipage}\hfill
    \begin{minipage}{0.5\linewidth}
        \centering
        \includegraphics[width=\linewidth]{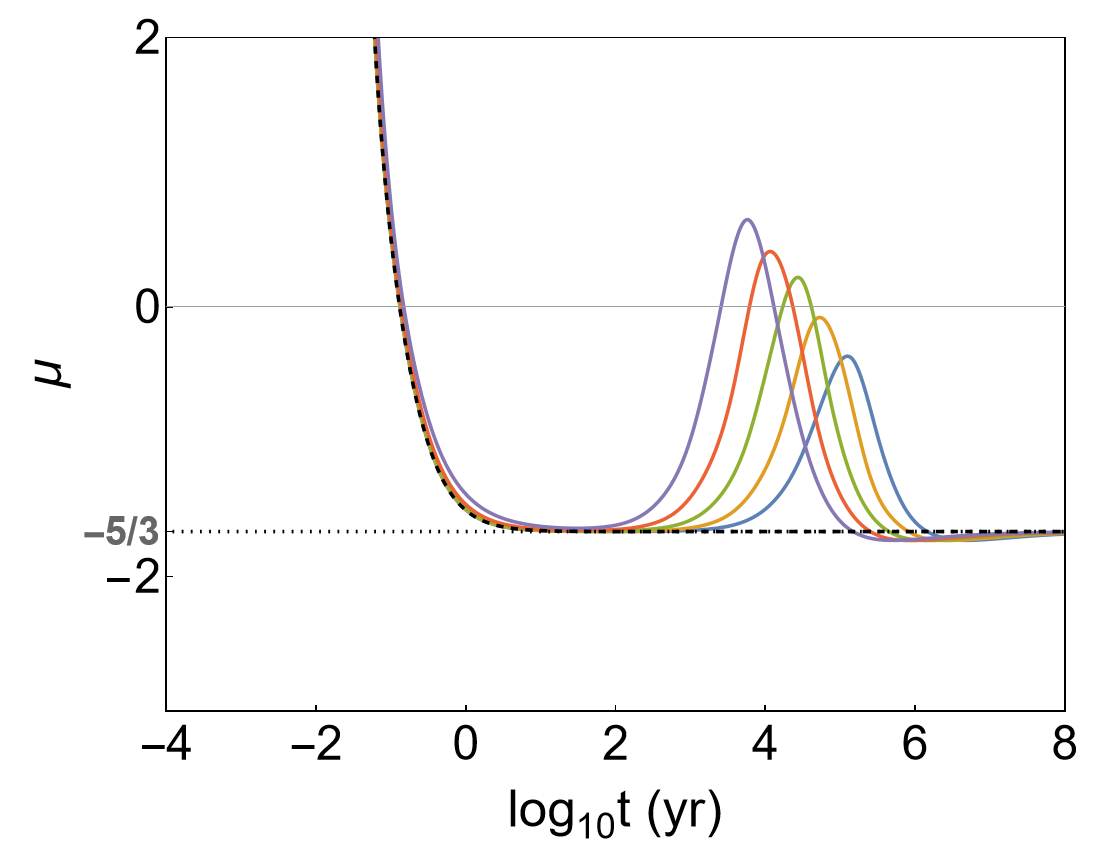}
    \end{minipage}
    \caption{Left: Evolution of the fallback accretion rate for different choices of the galactic mass. For comparison, the purely Keplerian prediction is shown as a black dashed curve. In all cases, the black hole mass is fixed at $10^4 \, M_{\odot}$ with a penetration factor $\delta = 1$, while the surrounding Hernquist potential is characterized by a constant scale length of $10^8 \, R_{\odot}$. The faint dotted line in the background indicates the canonical $t^{-5/3}$ fallback scaling. Right: Time evolution of the corresponding instantaneous power-law slope. The dotted horizontal line marks the standard $-5/3$ decay, while the solid horizontal line at zero separates regimes where the light curve exhibits a temporary rise (positive slope) from those where the decay simply becomes shallower than the canonical behavior.}
    \label{fig:galmass}
\end{figure*}

As shown in Figure \ref{fig:galmass}, the galactic mass indeed controls both the prominence and temporal onset of these signatures. In lower-mass environments, the light curves only display a modest deviation from the canonical decay, often in the form of a broad shallow tail that appears late in the evolution. Meanwhile, for heavier hosts, the deviations are not only more pronounced but also emerge a bit earlier, allowing the double-peaked structure to stand out more clearly.

The evolution of the power-law index further highlights the role of the galactic mass. As we increase the mass of the environment, the index shifts toward less negative values, reflecting slower declines in the accretion rate. In sufficiently massive environments, the index can even flip positive for a while, signaling a rebrightening episode. Overall, adding more galactic mass not only amplifies the accretion levels of these late-time features but also brings them closer to the initial disruption.

Though less apparent, the host potential also subtly affects the initial peak. This is not surprising because the total amount of stellar debris is fixed in the system. So, if a fraction of that material is delayed by the external potential, the initial fallback will be substantially diminished, resulting in a suppressed initial peak. This construes a second, albeit minor, influence of the galactic mass, for it can both alter the late-time behavior of the light curve, and affect the early-time energetics of the flare as well.

\subsubsection{Across varying scale lengths} \label{sec:sl}

Now, we examine how the light curves respond to variations in the scale length while keeping the galactic mass fixed. We also consider an extreme case involving a highly compact environment. Although such configurations may be physically unrealistic, or at least not supported by current observations, they provide a useful diagnostic for assessing the full extent of environmental influence on TDE evolution.

\begin{figure}
    \centering
    \begin{minipage}{0.5\linewidth}
        \centering
        \includegraphics[width=\linewidth]{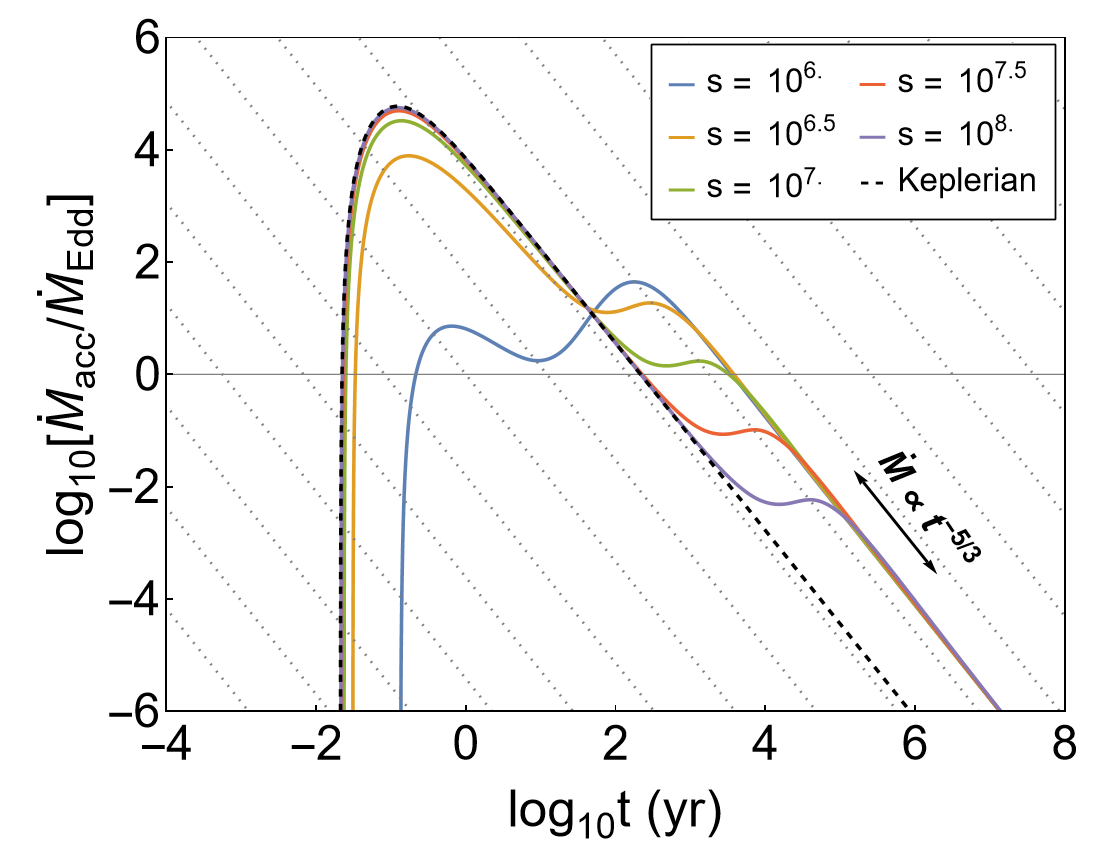}
    \end{minipage}\hfill
    \begin{minipage}{0.5\linewidth}
        \centering
        \includegraphics[width=\linewidth]{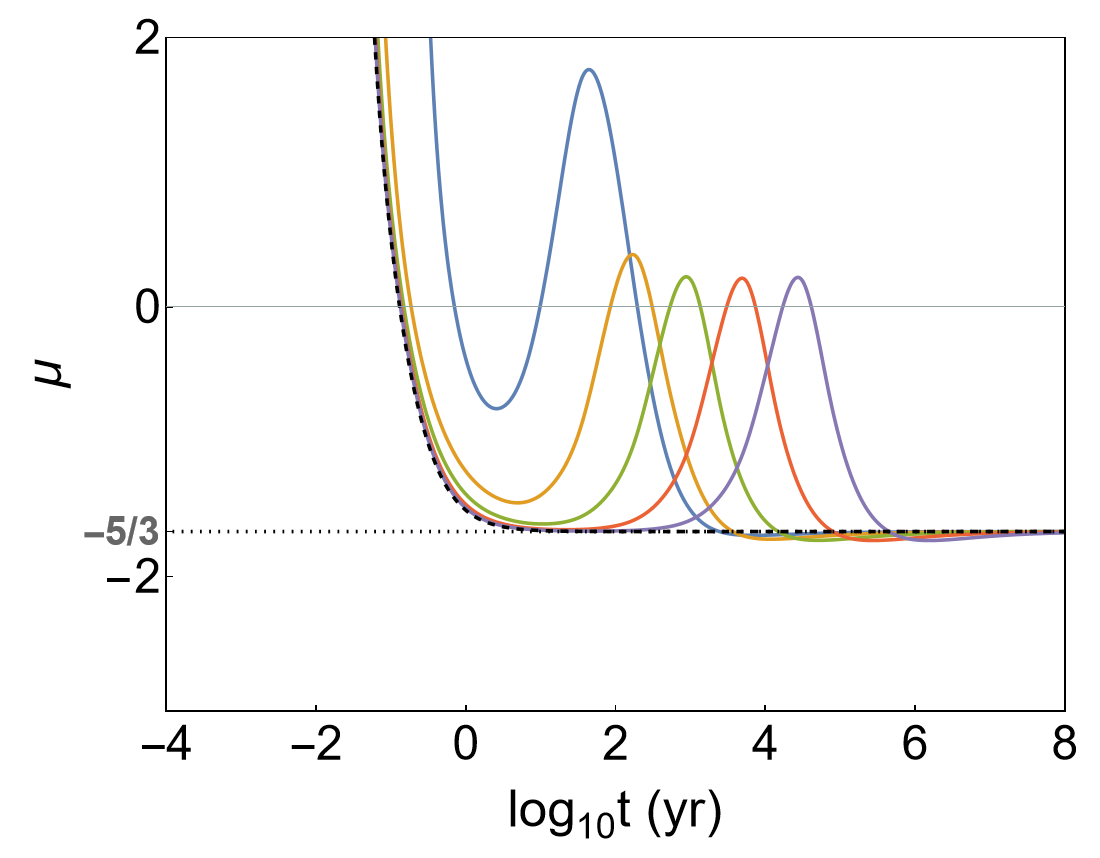}
    \end{minipage}
    \caption{Same setup as in Figure \ref{fig:galmass}, except that the galactic scale radius $s$ is varied while the total galactic mass is held fixed at $m_{\mathrm{gal}} = 10^7$.}
    \label{fig:sl}
\end{figure}

Figure \ref{fig:sl} (right panel) illustrates the time evolution of the mass accretion rate and the corresponding power-law index for different galactic scale lengths. As with increasing the host mass, varying the scale length also modifies both the strength and the timing of the environmental imprints. In contrast to the mass dependence, however, the response here is noticeably more sensitive to changes in $s$.

This sensitivity is most clearly exposed in the evolution of its power-law index. As the scale length increases, the deviations from the canonical Keplerian behavior are progressively shifted to later times and eventually become observationally irrelevant. Physically, a larger scale length spreads the same amount of mass over a much larger volume, weakening the gravitational field felt by the returning debris. In this regime, the black hole dominates the dynamics for an extended period, and the fallback naturally resembles that of an isolated system. This provides a simple explanation for why most observed TDEs appear as single, smooth outbursts. In typical galaxies, the stellar environment is sufficiently diffuse that its imprint on the fallback only emerges long after the flare has faded from view.

The picture changes qualitatively as the galactic potential becomes highly compact. Decreasing the scale length concentrates the mass closer to the black hole, steepening the local gravitational field experienced by the debris and allowing the environment to compete dynamically at much earlier times. In this regime, the secondary features identified earlier are no longer relegated to the distant tail of the light curve but can intrude into phases that would normally be considered part of the "early-time" evolution.

An extreme example is shown by the most compact model in Figure \ref{fig:sl} (blue curve). Here, the initial accretion peak is noticeably suppressed, while the subsequent rebrightening becomes both earlier and more pronounced, even exceeding the primary flare. Although such configurations may be rare or absent in real galaxies, they demonstrate, theoretically at least, that sufficiently concentrated environments can fundamentally reshape the fallback curve, producing strong, early rebrightening behavior that departs sharply from standard Keplerian expectations.

\subsection{Effects of encounter depth} \label{sec:pf}

\begin{figure*}
    \centering
    \begin{minipage}{0.5\linewidth}
        \centering
        \includegraphics[width=\linewidth]{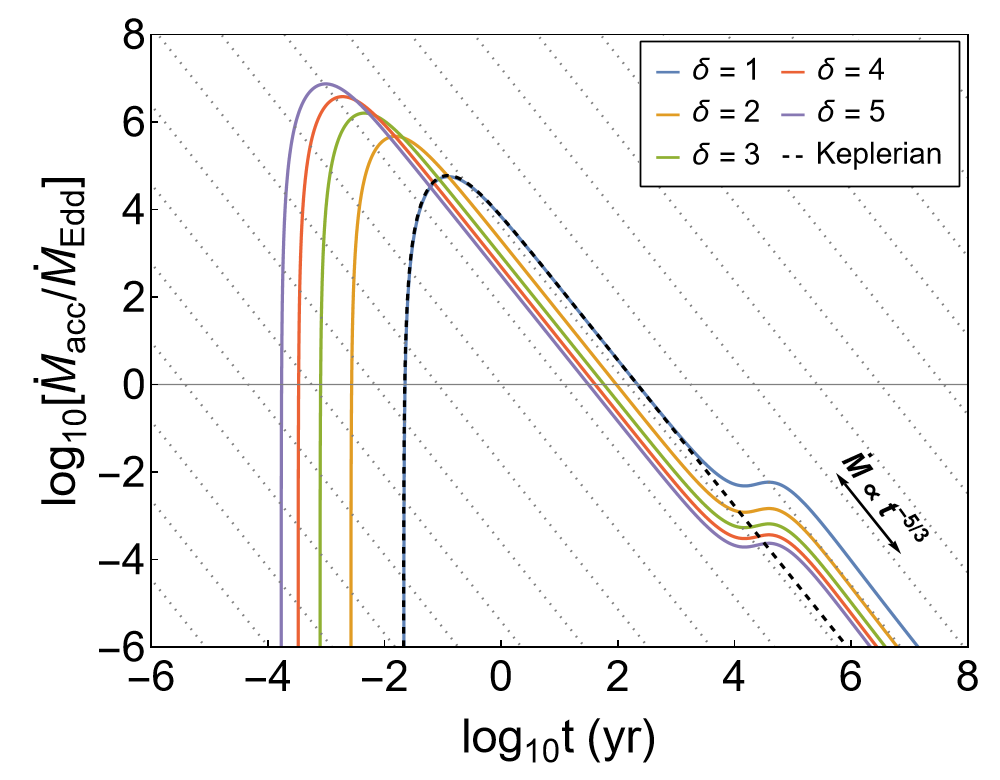}
    \end{minipage}\hfill
    \begin{minipage}{0.5\linewidth}
        \centering
        \includegraphics[width=\linewidth]{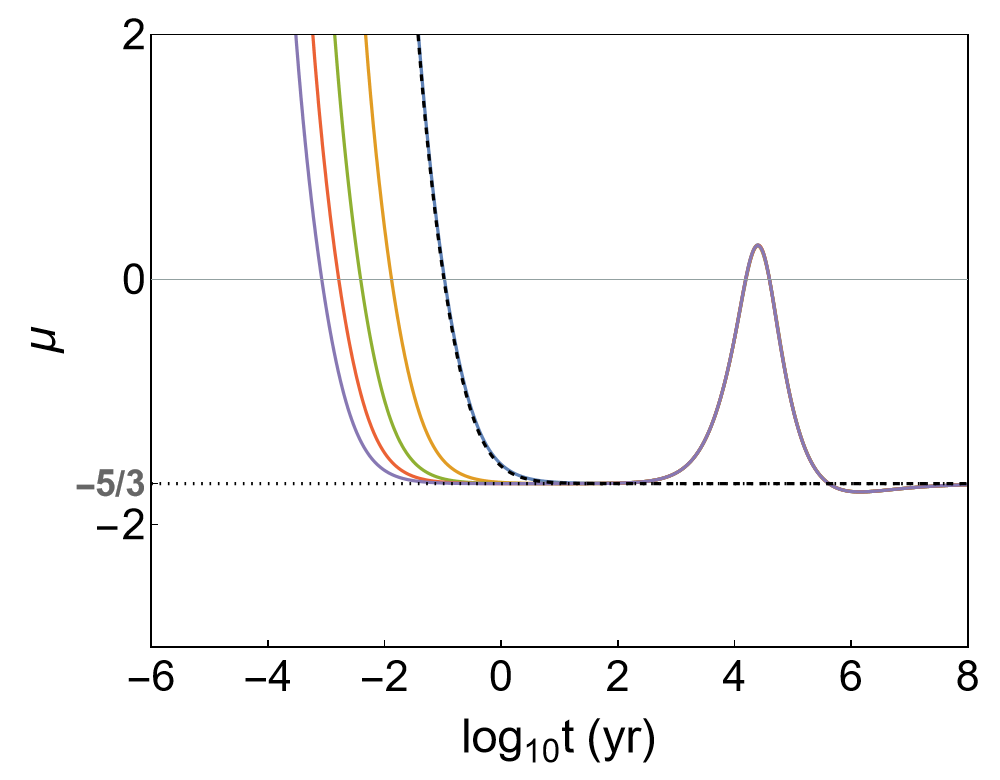}
    \end{minipage}
    \caption{Left: Time evolution of the mass accretion rate for different encounter depths, parameterized by the penetration factor $\delta$. The black dashed curve represents the standard Keplerian prediction. In all models presented here, the black hole mass is fixed at $10^4 \, M_{\odot}$, while the surrounding Hernquist potential has a total mass of $10^7 \, M_{\odot}$ and a scale length of $10^8 \, R_{\odot}$. The dotted lines in the background marks the canonical $t^{-5/3}$ fallback scaling. Right: Evolution of the corresponding instantaneous power-law slope. The dotted and solid horizontal reference lines indicate the same fiducial slopes used in the previous figures.}
    \label{fig:pf}
\end{figure*}

We now examine the role of the encounter depth and how it modifies the predicted light curves. The results are shown in Figure \ref{fig:pf}, where the accretion curves are generated for several values of the penetration factor $\delta$. 

Guided by the limits derived in Appendix \ref{sec:appB}, we restrict our analysis to $1 \leq \delta \leq 5$. Although Table \ref{tab:delta} shows that for a black hole mass $m_{\mathrm{bh}} = 10^4\,M_{\odot}$ the formal Newtonian upper bound on $\delta$ can be as large as $100$ or more, such extreme encounters would place the star very close to the capture radius, where relativistic effects become increasingly important. In this regime, the Newtonian treatment adopted in this work may no longer provide reliable results.

As a conservative choice, we therefore require that
\begin{equation}
    R_{\mathrm{p}} \gtrsim 100\,R_{\mathrm{S}}.
\end{equation}
For a $10^4\,M_{\odot}$ black hole, the Schwarzschild radius is $R_{\mathrm{S}} \approx 0.04\,R_{\odot}$, so this condition corresponds to $R_{\mathrm{p}} \gtrsim 4\,R_{\odot}$.
Since the tidal radius for this system is $R_{\mathrm{t}} \approx 21.5\,R_{\odot}$, the choice $\delta = 5$ yields $R_{\mathrm{p}} \approx 4.3\,R_{\odot}$, which safely satisfies the above constraint.

The left panel of Figure \ref{fig:pf} shows that the penetration factor directly dictates the timing of the flare. As $\delta$ increases, the peak of the light curve moves earlier along the timeline. This behavior is expected because a deeper encounter places the disrupted star closer to the black hole, which shortens the orbital periods of the debris and, therefore, reduces the fallback timescale. This also explains why the accretion peak appears earlier compared to more grazing encounters.

Moreover, the penetration depth also influences the amplitude of the secondary peaks appearing in the light curve at much later times. In particular, the secondary features emerge at different accretion levels depending on $\delta$. Models with larger penetration factors tend to display these signatures at lower fallback rates. To explain this behavior, recall that the total debris mass available to fall back is fixed. Since deeper encounters trigger the accretion phase earlier, a larger fraction of the debris has already returned to the black hole by the time the environmental effects begin to manifest. Therefore, less material remains available to power the secondary peaks, making them somewhat suppressed for higher $\delta$ compared to shallower encounters. This can also be understood as a consequence of the stronger dominance of the black hole’s gravity at smaller pericenter distances in high-$\delta$ encounters, which reduces the impact of environmental effects.

Interestingly, the same is not true for the slope evolution. The right panel of Figure \ref{fig:pf} shows that the instantaneous power-law index of the secondary features follows identical behaviors across the entire range of penetration factors. In other words, while the penetration depth clearly shifts the absolute accretion levels of the later peaks, it does not significantly alter how the slope itself evolves. The deviations from the canonical slope scaling, therefore, appear to be largely insensitive to the specific choice of $\delta$.


\section{Parameter Study} \label{sec:par}

In this section, we move beyond individual examples and systematically chart the regions of parameter space where departures from purely Keplerian behavior, such as flattening of the light curve or the appearance of secondary luminosity peaks occur. To do so, we explore a broader and more nuanced range of host properties, including configurations that may lie at the edges, or even outside of observationally supported regimes. While some of these cases may be rare or extreme, they are still essential for mapping the full theoretical landscape.

\subsection{Shallow slopes and rebrightenings: A map of the parameter space}

Recall that the model depends on four main parameters. Instead of attempting to visualize the full 4d space all at once, we examine 2d slices that isolate the dependence on the environmental parameters. We focus on the plane defined by $(s, m_{\mathrm{gal}})$ and generate these slices under different disruption conditions. Specifically, we consider three representative black hole masses ($m_{\mathrm{bh}} = 10^4$, $10^5, 10^6$) and encounter depths ($\delta = 2,3,4$), producing a total of six distinct parameter slices through the larger 4d space. Each slice corresponds to a particular disruption setup defined by a given black hole mass and encounter depth, while the environmental parameters $(s, m_{\mathrm{gal}})$ are allowed to vary freely across the grid. The host environment is still described with a Hernquist profile, just as in the previous section.

For every slice, we construct a $200 \times 200$ grid in $(s, m_{\mathrm{gal}})$, which ends up producing about $40,000$ synthetic light curves. Both $s$ and $m_{\mathrm{gal}}$ are spaced logarithmically between $10^4$ and $10^{12}$. Admittedly, not every point in this grid should be interpreted as a realistic galaxy configuration. Nevertheless, allowing the grid to stretch this far exposes the full extent of how the environmental structure can, in principle, influence TDE light curves.

The light curves are classified into three different classes based on their evolution. First are light curves with phases of late-time shallower-than-canonical decays. In these cases, the slope remains negative, but drifts away from the familiar $t^{-5/3}$ scaling. The second class of models develops rebrightening episodes. After the initial decline, the luminosity rises again and forms a secondary peak, which shows up in the slope evolution as a brief positive slope index. The remaining cases behave more conventionally, following a smooth, monotonic Keplerian-like decay without any secondary structure.

For models that do show environmental signatures, whether in the form of late-time flattening or rebrightening, we track two key quantities. First, we record the time at which the feature begins to appear. Second, we measure how luminous the system is at that moment by comparing the accretion rate to the Eddington limit.


\subsubsection{Temporal onset of environmental signatures}

\begin{figure}
    \centering
    \begin{minipage}{0.33\linewidth}
        \centering
        \includegraphics[width=\linewidth]{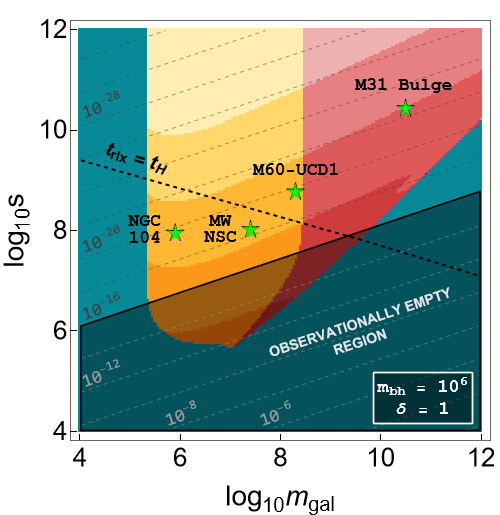}
    \end{minipage}\hfill
    \begin{minipage}{0.33\linewidth}
        \centering
        \includegraphics[width=\linewidth]{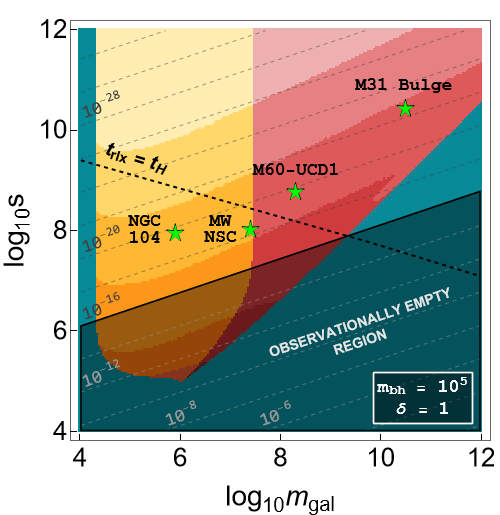}
    \end{minipage}\hfill
    \begin{minipage}{0.33\linewidth}
        \centering
        \includegraphics[width=\linewidth]{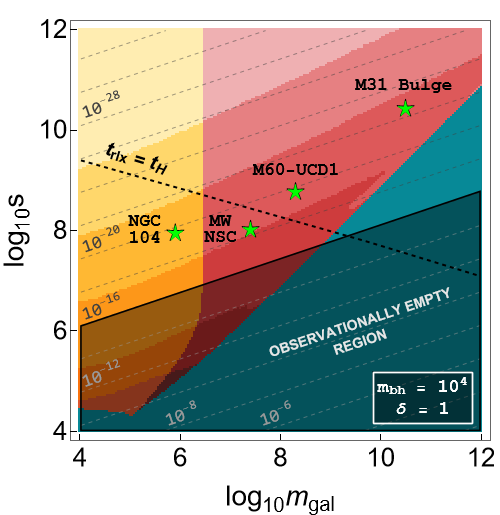}
    \end{minipage}
    \begin{minipage}{\linewidth}
        \centering
        \includegraphics[width=0.5\linewidth]{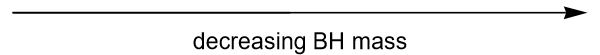}
    \end{minipage}\hfill
    \begin{minipage}{0.33\linewidth}
        \centering
        \includegraphics[width=\linewidth]{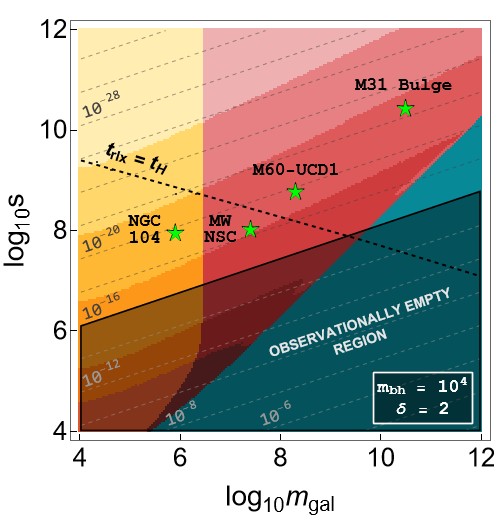}
    \end{minipage}\hfill
    \begin{minipage}{0.33\linewidth}
        \centering
        \includegraphics[width=\linewidth]{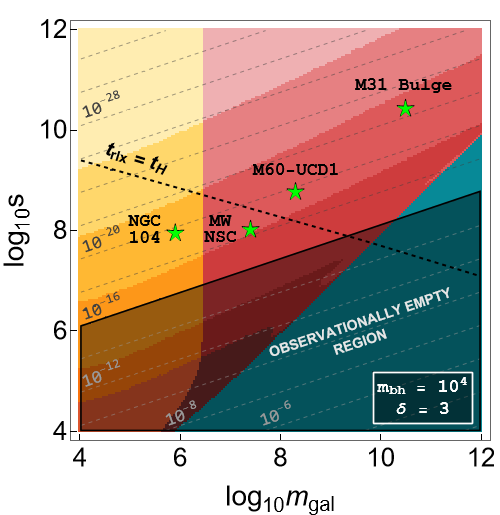}
    \end{minipage}\hfill
    \begin{minipage}{0.33\linewidth}
        \centering
        \includegraphics[width=\linewidth]{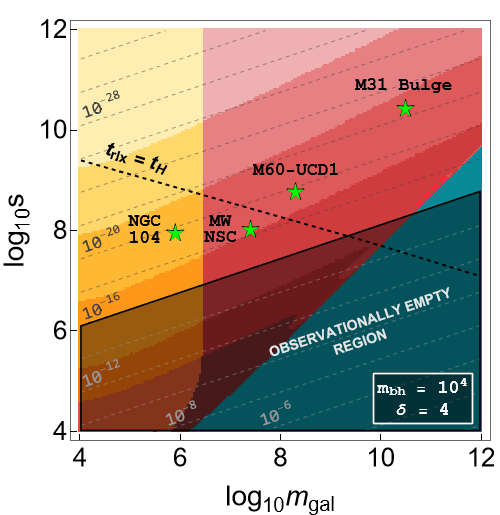}
    \end{minipage}\hfill
    \begin{minipage}{\linewidth}
        \centering
        \includegraphics[width=0.5\linewidth]{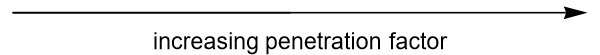}
        \vspace{10pt}
    \end{minipage}
    \setlength{\fboxsep}{6pt}
    \fbox{\includegraphics[width=0.9\linewidth]{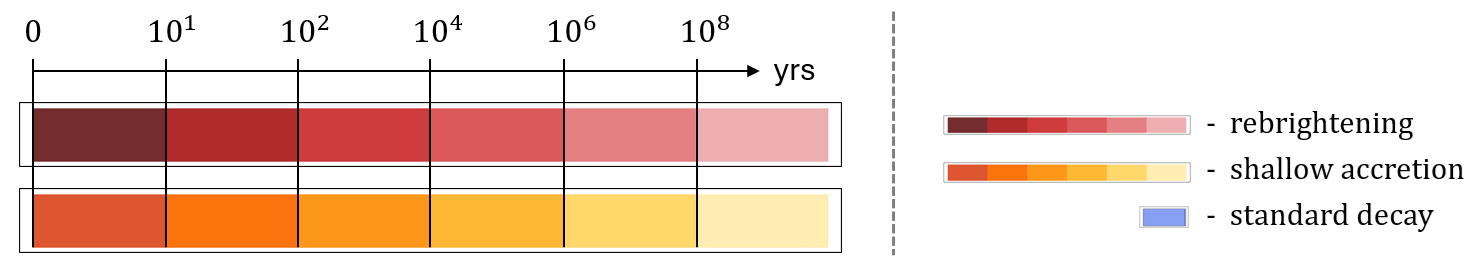}}
    
    \caption{Temporal locations of environmental signatures in TDE light curves. Regions producing rebrightening episodes are shown in shades of red, shallower-than-standard late-time decays in shades of yellow, and standard Keplerian-like light curves in blue. Color intensity indicates the onset time of the environmental feature: darker shades correspond to earlier appearance, while lighter shades indicate progressively later times, spaced roughly by one to two orders of magnitude in years. The timing legend is shown in the boxed panel below for reference. Diagonal contours trace constant mean galactic density. The dashed black curve marks the locus where the two-body relaxation time equals the Hubble time ($t_{\mathrm{rlx}}=t_H$). The dark shaded area represent observationally empty regions of the parameter space, while green stars denote representative observed galactic environments. Upper panels: black hole mass decreases from left to right, penetration factor fixed at $\delta=1$. Lower panels: penetration factor increases from left to right, black hole mass fixed at $m_{\mathrm{bh}}=10^4$.}
    \label{fig:timing}
\end{figure}

Figure \ref{fig:timing} summarizes when environmental effects start to leave visible marks on the fallback evolution. Each panel explores the $(s, m_{\mathrm{gal}})$ parameter space for a particular choice of black hole mass and penetration factor. The color scheme separates the different kinds of behaviors that emerge in the light curves. Red regions correspond to models that produce a secondary brightening. Yellow regions correspond to light curves that does not develop a second peak but instead settles into a shallower late-time decay. Blue regions represent cases in which no noticeable environmental imprint emerges within the duration of the simulation.

To provide some observational context, several real stellar systems are overlayed on the diagram using green stars. Their reported half-light radii from the literature are first converted into Hernquist scale radii using the relation $r_{\mathrm{hl}}=1.815 b$. Of course, actual galaxies rarely follow an exact Hernquist profile, so this step should be viewed purely as a convenient approximation. The goal is simply to place a diverse set of environments onto a common footing for comparison. The examples plotted here include the Milky Way nuclear star cluster (MW NSC) \citep{bc}, the globular cluster NGC 104 \citep{bd, be, bf}, the ultra-compact dwarf M60-UCD1 \citep{bg}, and the bulge component of M31 \citep{bh}. Interestingly, the M31 bulge has, in fact, been modeled directly with a Hernquist profile by Geehan \citep{bi}, who obtained a scale radius of approximately $b \sim 0.61 \,\mathrm{kpc}$.

One of the first things that becomes apparent from the figure is that real systems cluster within only a relatively small part of the diagram. In fact, they cluster within similar density strips. Much of the available parameter space below them remains unoccupied, shown as the dark shaded area and is labeled “observationally empty". It is important not to interpret this boundary too literally. It does not represent a fundamental physical limit on galaxy properties; it simply reflects the current state of observations. At present, we do not know of stellar systems that fall within those parts of the diagram.

To make the diagram easier to interpret, we also plot contours of constant stellar density in the $(s, m_{\mathrm{gal}})$ plane. These contours correspond to the average density evaluated at the scale radius. This is calculated by evaluating the density profiles listed in Table \ref{tab:pot} at the scale radius $b$ then normalizing using Eq. \eqref{eq:fidrho}. The resulting contours run diagonally across the diagram, with densities increasing as one moves from the upper-left toward the lower-right corner.

If we line up the density contours against the colored regions, we see that environments with similar densities tend to fall within the same timing regimes. This indicates that the relevant controlling parameter is not the scale radius or the total mass individually, but rather their combination through the effective compactness of the mass distribution.

For the real galactic settings considered here, their environmental effects on the light curves generally appear only at late stages of the disruption as green stars position themselves within the $10^4-10^6$ time range. Regions of the diagram that would produce very early environmental signatures lie almost entirely within the observationally empty zone. However, within the range of known systems and the parameters explored here, those regimes are simply not realized. This provides a natural explanation for why standard Keplerian fallback models often reproduce the early phases of observed TDE light curves quite successfully. In most realistic environments, the external potential does not begin to matter until the event has already faded substantially at later times.

The dependence on black hole mass introduces an additional layer to this picture (see upper panels). Lowering the black hole mass expands the portion of parameter space that produces either rebrightening or shallow late-time decay. As a consequence, the MW NSC, for example, is expected to imprint a shallow late-time tail in SMBH systems, but can produce a rebrightening phase when the central object is an IMBH. In other words, the dynamical importance of the host environment increases as the central black hole becomes less massive. This is consistent with what we have seen in Section \ref{sec:bhmass}.

Compared with the black hole mass, the penetration factor has a noticeably weaker influence on the overall structure of the diagram. At first glance, the lower panels look almost unchanged as the penetration factor varies. Only upon closer inspection do small differences become noticeable. Increasing the penetration factor slightly expands the regions associated with secondary features. In effect, some of the early-time signatures begin to extend towards the observable region of the parameter space. Extrapolating this trend suggests that, for sufficiently deep encounters, environmental signatures could potentially emerge on observable timescales for realistic galactic environments. However, pushing the penetration factor to very high values introduces an important caveat. Deeply penetrating events bring the stellar debris much closer to the black hole, where relativistic effects become increasingly significant. Since our model does not explicitly include these corrections, its assumptions may begin to break down in that regime. For this reason, the behavior seen at large penetration factors should be interpreted cautiously.

Overall, both the black hole mass and the penetration factor still play a role in determining when environmental signatures begin to appear. The black hole mass remains the dominant factor, strongly shifting the timing of these features.


\subsubsection{\label{sec:luminosity}Eddington regimes of the resulting accretion}

\begin{figure}
    \centering
    \begin{minipage}{0.33\linewidth}
        \centering
        \includegraphics[width=\linewidth]{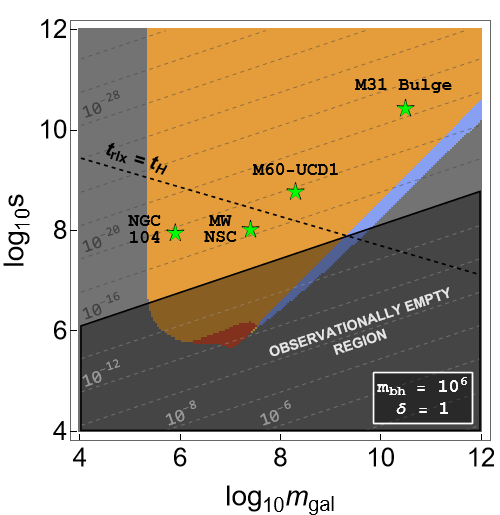}
    \end{minipage}\hfill
    \begin{minipage}{0.33\linewidth}
        \centering
        \includegraphics[width=\linewidth]{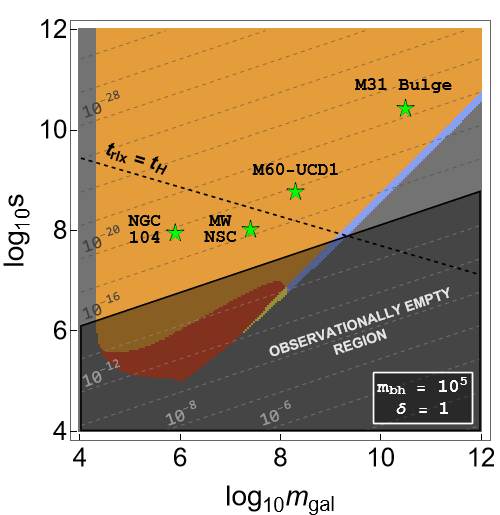}
    \end{minipage}\hfill
    \begin{minipage}{0.33\linewidth}
        \centering
        \includegraphics[width=\linewidth]{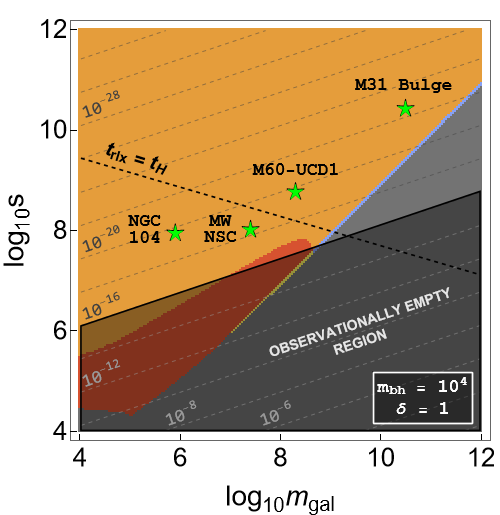}
    \end{minipage}
    \begin{minipage}{\linewidth}
        \centering
        \includegraphics[width=0.5\linewidth]{Figures/bh_arrow.png}
    \end{minipage}\hfill
    \begin{minipage}{0.33\linewidth}
        \centering
        \includegraphics[width=\linewidth]{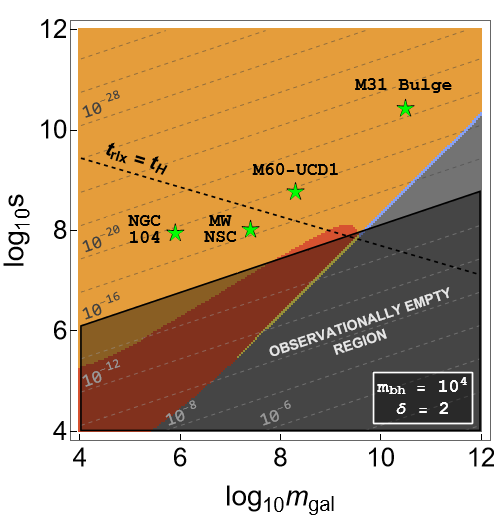}
    \end{minipage}\hfill
    \begin{minipage}{0.33\linewidth}
        \centering
        \includegraphics[width=\linewidth]{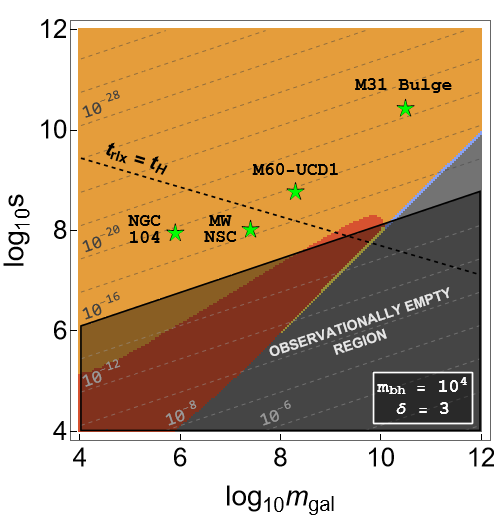}
    \end{minipage}\hfill
    \begin{minipage}{0.33\linewidth}
        \centering
        \includegraphics[width=\linewidth]{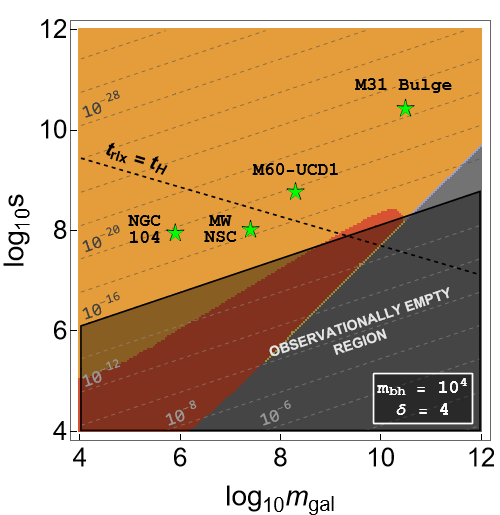}
    \end{minipage}\hfill
    \begin{minipage}{\linewidth}
        \centering
        \includegraphics[width=0.5\linewidth]
        {Figures/pf_arrow.png}
        \vspace{10pt}
    \end{minipage} \hfill
    \setlength{\fboxsep}{6pt}
    \fbox{\includegraphics[width=0.6\linewidth]{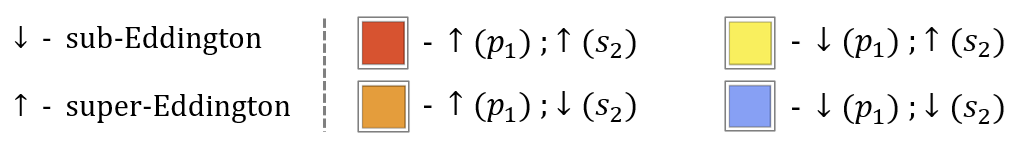}}
    
   \caption{Parameter space highlighting only regions where secondary environmental features appear. Areas without such features are shown in grayscale for contrast. Colors indicate the accretion level relative to the Eddington limit: red for both the primary peak ($p_1$) and the secondary feature ($s_2$) being super-Eddington, orange for $p_1$ super-Eddington but $s_2$ sub-Eddington, yellow for $p_1$ sub-Eddington but $s_2$ super-Eddington, and blue for both sub-Eddington. A luminosity legend is included for reference. Additional markers and overlays are exactly similar as Figure \ref{fig:timing}.}
    \label{fig:luminosity}
\end{figure}

Figure \ref{fig:luminosity} focuses on the luminosity regimes of the secondary features across the same $(s, m_{\mathrm{gal}})$ planes. Here, we explore how bright these signatures are relative to the Eddington limit, and therefore how observable they might be. The colors indicate the Eddington status of both the primary fallback peak and any secondary signature produced by the galactic environment.

Notice in the plots that most of the populated parameter space is orange. In other words, for environments resembling known galaxies, the initial TDE flare is typically super-Eddington, while the later environmental features are not. Although the galactic potential can reshape the late-time fallback behavior, the resulting signatures generally occur at luminosities below the Eddington limit. This means that shallow late-time tails or delayed rebrightening phases are expected to be fainter than the original peak and are unlikely to drive strong outflows or powerful jet activity.

Situations where the secondary feature itself becomes super-Eddington are comparatively rare. The yellow and red regions occupy only narrow regions of the diagram and are concentrated toward environments with extremely high compactness. In these cases, the debris stream is perturbed strongly enough that the late-time accretion episode approaches, or even exceeds, the Eddington limit. However, notice that these regions fall largely inside the observationally empty part of the parameter space, implying that such extreme behavior is unlikely to occur in the types of galactic environments currently known. The locations of the representative systems reinforce this general picture. The MW NSC, NGC 104, M60-UCD1, and the M31 bulge all lie comfortably within the orange region. For these environments, the primary TDE flare should be super-Eddington, but any environmental imprint on the light curve is expected to remain sub-Eddington and therefore relatively subtle observationally.

Black hole mass again plays an important role here. As the black hole mass decreases, the red regions broaden. Lower-mass black holes provide more favorable conditions for energetically significant environmental signatures. For example, in the $m_{\mathrm{bh}}=10^4$ case (upper-rightmost panel), a small red patch begin to appear out of the observationally empty region, hinting that IMBHs, and in the extreme limit even stellar-mass black holes, could potentially host systems where the environmental rebrightening becomes genuinely luminous.

The penetration factor has a more subtle effect. The peak accretion levels do not change dramatically with its value, but a closer look reveals that increasing the penetration factor slightly enlarges the regions of the parameter space with environmental signatures. Still, the resulting environmental imprints rarely compete with the brightness of the initial fallback peak, except in extreme regions of parameter space that are not currently populated by observed galactic systems.


\subsubsection{Dynamical stability of the host environment}

Now, we discuss the $t_{\mathrm{rlx}} = t_H$ dashed line appearing in the panels of Figures \ref{fig:timing} and \ref{fig:luminosity}. This is the locus of $(s, m_{\mathrm{gal}})$ combinations for which the two-body relaxation time $t_{\mathrm{rlx}}$ equals the Hubble time $t_H$. The curve serves as a useful reference for determining whether a stellar system can reasonably be treated as dynamically frozen over cosmological timescales or whether internal relaxation processes are likely to reshape it over time. For reference, the local relaxation time is given by \citep{bj}
\begin{align}
    t_{\mathrm{rlx}} = \frac{0.34 \sigma(r)^3}{\rho(r) m_\star G^2 \ln \Lambda},
\end{align}
where $\sigma(r)$ is the one-dimensional stellar velocity dispersion and $\ln \Lambda$ is the Coulomb logarithm. In our dimensionless units, this expression can be written more conveniently as
\begin{align}
    \tilde{t}_{\mathrm{rlx}} \equiv \frac{t_{\mathrm{rlx}}}{t_0} \simeq 0.34 \left(\frac{\sqrt{m_{\mathrm{enc}}(<s)}}{\ln |m_{\mathrm{enc}}(<s)|}\right)s^{3/2},
\end{align}
with $t_0 = \sqrt{R_\odot^3 / G M_\odot}$, $\ln \Lambda \simeq \ln |m_{\mathrm{enc}}(<s)|$, and $m_{\mathrm{enc}}(<s)$ representing the enclosed mass within the scale radius $s$.

This relaxation boundary effectively separates the parameter space into two different dynamical regimes. Systems lying above the curve have $t_{\mathrm{rlx}} \gtrsim t_H$ and therefore evolve only slowly through stellar encounters. Within the timescales relevant for our simulations, they can be treated as essentially static backgrounds. Systems below the curve, on the other hand, have $t_{\mathrm{rlx}} \lesssim t_H$ and are expected to undergo measurable dynamical evolution over cosmic time. Comparing the relaxation time to the predicted onset time of environmental signatures, $t_{\mathrm{sig}}$, helps clarify whether those features develop in a quasi-static stellar potential or in one that may itself be evolving.

The locations of the considered stellar systems follow these expectations quite well. The MW NSC, which is representative of NSCs more generally, falls below the relaxation boundary. This is consistent with estimates of its half-mass relaxation time of roughly $11$ Gyr \citep{bk}, and with the broader understanding that two-body relaxation can significantly reshape NSCs over $\sim10$ Gyr \citep{bl}. Even so, their relatively large masses and sizes mean that NSCs evolve more slowly than typical globular clusters \citep{bm}. Indeed, for most globular clusters the median relaxation time is closer to $10^9$ years \citep{ai}, placing them well within the collisional regime.

Ultra-compact dwarfs behave differently. Systems such as M60-UCD1 are expected to have relaxation times that exceed the Hubble time, meaning they behave largely collisionlessly on these scales. UCDs occupy an intermediate category between massive star clusters and small galaxies, and their median relaxation times often surpass the age of the Universe. This is precisely why they are commonly described as collisionless stellar systems on cosmological timescales \citep{bn, bo}.

This distinction between collisional and collisionless environments matters for tidal disruption events. In collisional systems, frequent stellar encounters can perturb the debris stream, alter the fallback process, and in some cases interfere with the accretion flow onto the black hole. In collisionless systems, the background potential remains comparatively static, so the evolution of the debris is governed primarily by the large-scale gravitational field rather than by local stellar interactions.


\subsubsection{Behavior in limiting cases}

We refer again to Figure \ref{fig:timing} and examine the far edges of the parameter space. Consider first the case where the scale length $s$ becomes extremely small. In that limit, the surrounding stellar distribution collapses into a very compact configuration, effectively behaving like an additional point mass at the center. Dynamically, the debris then evolves almost exactly as it would in a purely Keplerian potential, except that the central mass is slightly more massive because the galactic mass simply adds to the black hole. The fallback curves, therefore, revert to the familiar Keplerian form with no noticeable late-time structure. This also explains why no secondary features appear at very small values of $s$. In that lower-most part of the parameter space, though not fully captured in the plots, the expectation is that these regions would be predominantly blue, indicating no noticeable environmental imprint. This limit mainly serves as a sanity check for the model rather than a realistic astrophysical scenario, since observed galaxies typically have scale lengths spanning from a few parsecs up to several kiloparsecs \citep{bp}.

On the other hand, when the scale length becomes very large, the same total galactic mass is distributed over a vast region that its gravitational pull near the disruption site becomes extremely weak. The debris stream then evolves almost as if the galaxy were not there at all, again producing fallback curves that closely resemble the purely isolated Keplerian case. This trend can be seen in Figure \ref{fig:timing}, although not immediately obvious since the regions remain colored, indicating that environmental effects are still present. But, take note that as $s$ increases, the onset of these environmental signatures steadily shifts to later and later times. Therefore, we can simply treat them as essentially Keplerian over the bulk of their evolution. Once $s \gtrsim 10^6$, the appearance of flattening or secondary brightening is postponed by more than a century after the initial disruption. This offers a simple explanation for why the majority of observed TDE light curves appear single-peaked and Keplerian-like within the typical observation windows of our surveys. Environmental traces may indeed be present, but they emerge only after the event has faded significantly, at times when long-term monitoring becomes extremely difficult. Even the densest environments we considered here are expected to produce detectable effects only at very late stages.

Another limiting behavior is when the galactic mass itself becomes very small. If $m_{\mathrm{gal}}$ is sufficiently low, the background potential contributes only a negligible perturbation to the debris orbits. As $m_{\mathrm{gal}} \rightarrow 0$, the surrounding galaxy effectively disappears from the problem, leaving behind the standard Keplerian picture in which the fallback dynamics are governed solely by the central black hole. We can safely expect the far-left side of Figure \ref{fig:timing} to be predominantly blue. Even though the regime with $m_{\mathrm{gal}} < 10^4$ is not explicitly included in the figure, the trend strongly suggests that it would fall into the region where no noticeable environmental signatures emerge.


\subsection{Beyond the Hernquist model: Environmental imprints in alternative galactic potentials}

\begin{figure}
    \centering
    \begin{minipage}{0.33\linewidth}
        \centering
        \includegraphics[width=\linewidth]{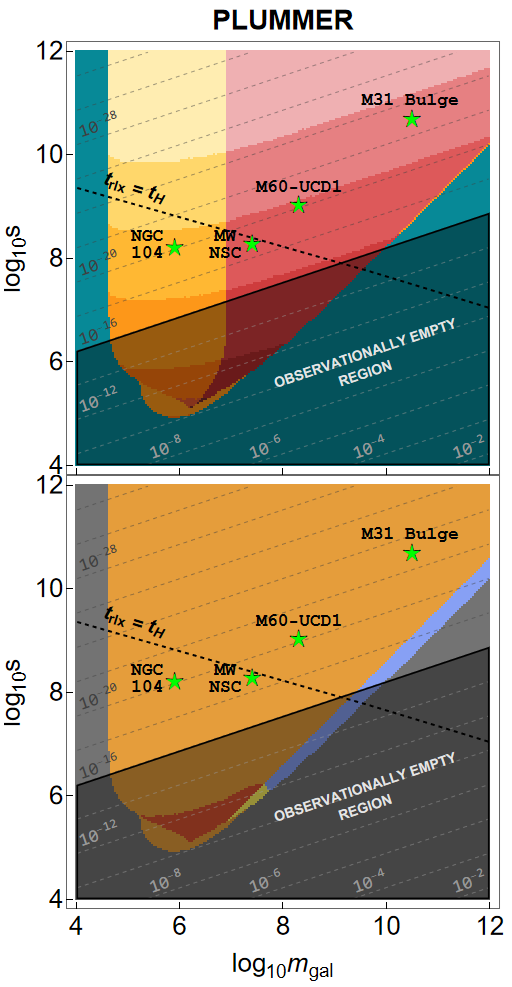}
        \vspace{2pt}
    \end{minipage}\hfill
    \begin{minipage}{0.33\linewidth}
        \centering
        \includegraphics[width=\linewidth]{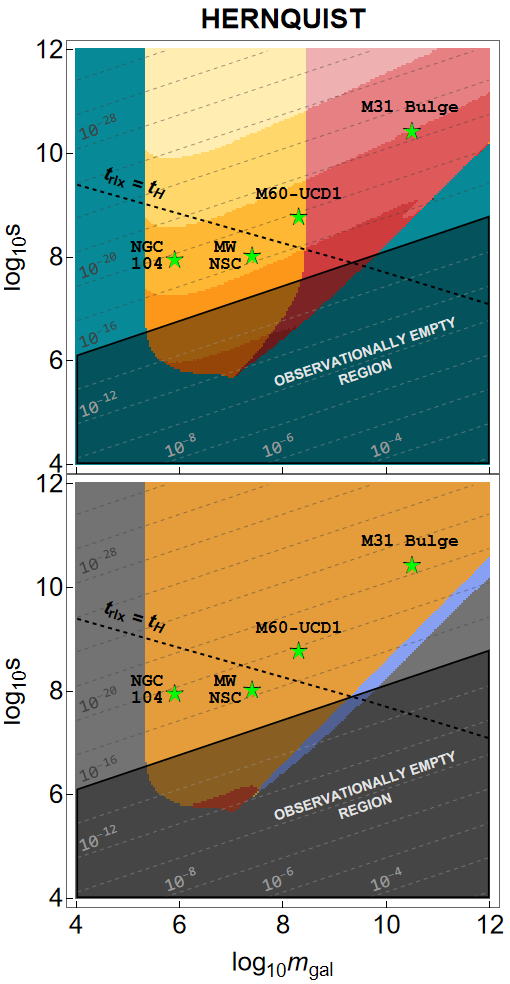}
        \vspace{2pt}
    \end{minipage}\hfill
    \begin{minipage}{0.33\linewidth}
        \centering
        \includegraphics[width=\linewidth]{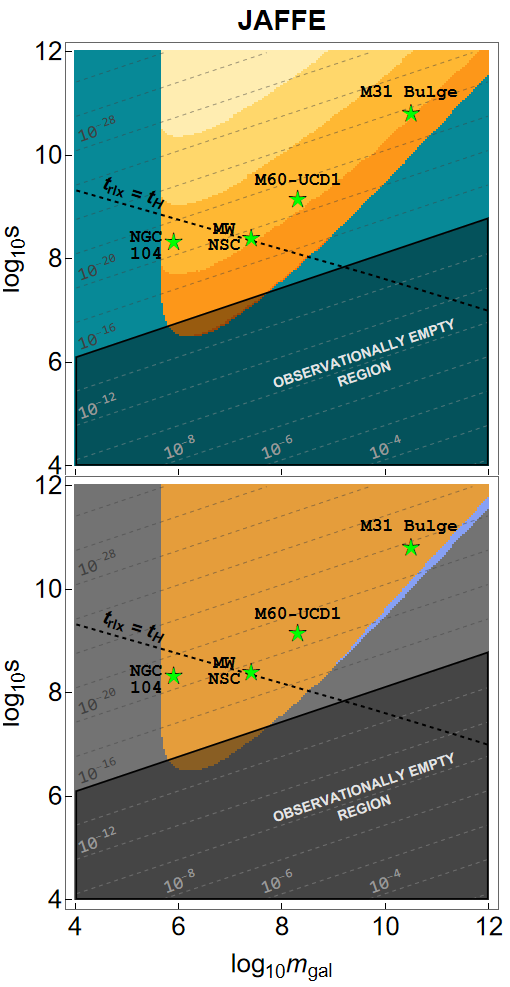}
        \vspace{2pt}
    \end{minipage}
    \fbox{\begin{minipage}{0.48\linewidth}
    \footnotesize\textbf{Upper panels:} Temporal onset
        \centering
        \includegraphics[width=\linewidth]{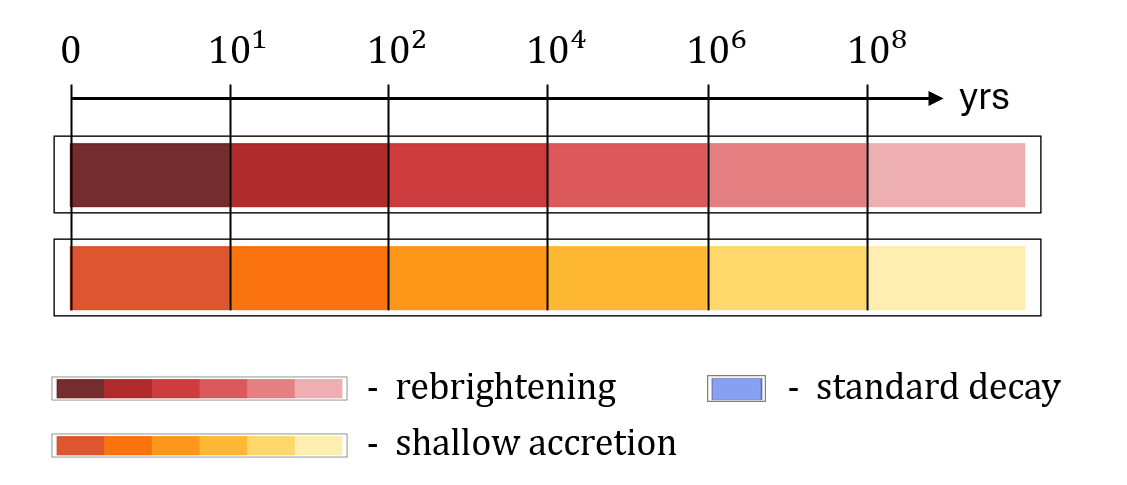}
    \end{minipage}}
    \hfill
    \fbox{\begin{minipage}{0.48\linewidth}
    \footnotesize\textbf{Lower panels:} Accretion regime
        \centering
        \includegraphics[width=\linewidth]{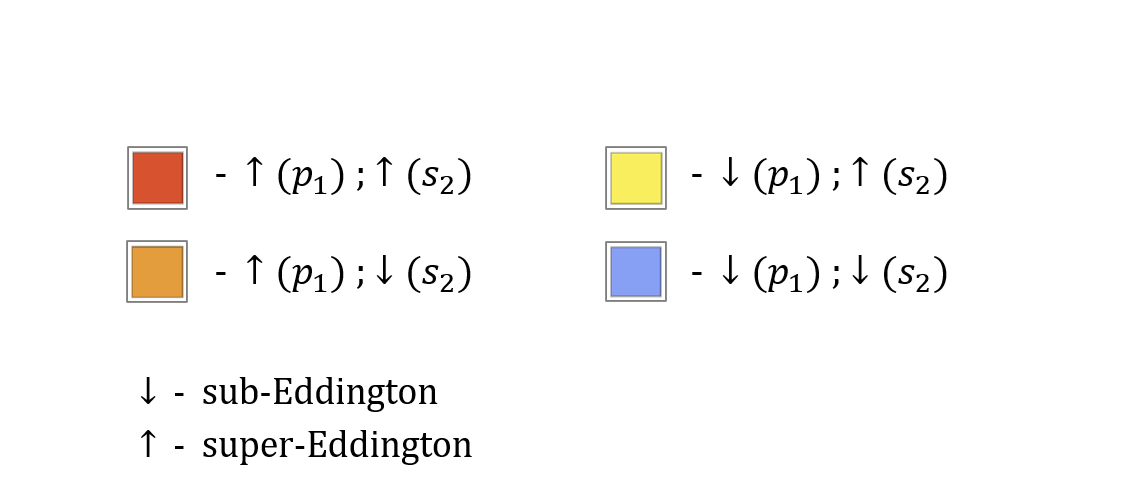}
    \end{minipage}}
    \caption{TDE light curve signatures for different galactic density profiles: Plummer (left), Hernquist (center), and Jaffe (right). Upper panels: Temporal onset of environmental signatures. Color intensity encodes the timing of secondary features, following the same convention and temporal legend as Figure \ref{fig:timing}. Lower panels: Corresponding accretion regimes of these environmental features, using the same color scheme as Figure \ref{fig:luminosity}. Dark-shaded regions mark observationally empty areas of parameter space, while green stars indicate representative observed systems mapped to each profile. The black dashed curve traces parameter combinations with relaxation times equal to the Hubble time. Background curves correspond to mean density contours evaluated at the scale radius of each model. All panels assume a fixed black hole mass of $m_{\mathrm{bh}} = 10^6$ and a penetration factor of $\delta = 1$.}
    \label{fig:profiles}
\end{figure}

So far, the surrounding galaxy has been modeled with a single density profile. While that is convenient for building intuition, real stellar systems are not all structured in the same way. Some galaxies have shallow cores, others develop steep cusps toward the center, and many fall somewhere in between. Because the spatial distribution of mass ultimately determines how strongly the debris stream feels the environment, it is natural to ask how sensitive the predicted light curves are to the choice of density profile itself. 

In this section, we repeat the same analysis using three widely used models for spherical stellar systems: the Plummer sphere, the Hernquist profile, and the Jaffe profile. The Plummer model represents a cored system, Hernquist describes a centrally concentrated but still relatively extended profile and the Jaffe, by contrast, is much more strongly cusped and packs a larger fraction of its mass closer to the center. The results of the simulations are summarized in Figure \ref{fig:profiles}. The upper panels map the times at which the secondary light curve features are expected to appear, using the same classification and color scheme introduced earlier. The lower panels, on the other hand, indicate the accretion levels of the predicted signatures relative to the Eddington limit.

One of the clearest differences shows up in the Jaffe models. For the black hole mass considered here, the parameter space does not produce any noticeable rebrightening episodes (the red regions that appeared in earlier figures). The fallback curves remain relatively smooth throughout their evolution. At most, they show mild flattening in the late-time decay, but they never develop the pronounced secondary peaks seen in other cases. A likely explanation has to do with how the total galactic mass is distributed. In a Jaffe profile, the mass is strongly concentrated toward the center and falls off rapidly outside the cusp. There simply is not enough material at larger radii to produce the kind of delayed dynamical response needed to generate a secondary peak. Even if one adjusts the scale radius or total mass, redistributing material outward without altering the fundamental steepness of the profile is difficult. As a result, the system tends to produce relatively featureless fallback curves.

The picture changes once the density profile becomes less centrally concentrated. Both the Hernquist and Plummer models produce rebrightening across noticeable portions of the parameter space. The Hernquist profile, in particular, sits in an interesting middle ground with a less cuspy profile. This leaves a non-negligible amount of stellar mass distributed at intermediate radii. That extended mass reservoir turns out to be important as this additional gravitational component can perturb the fallback in a way that delays more debris that powers the light curve at much later times.

The Plummer sphere, on the other hand, leads to a somewhat different behavior. Because this model has a finite-density core and a more diffuse central structure, more stellar mass are distributed at the outer regions of the system. This means the debris stream experiences the gravitational influence of the host environment more slowly and over longer timescales. In Figure \ref{fig:profiles}, the representative galaxies mapped under the Plummer assumption generally lie in color bands associated with later times. Take the M31 bulge as an example. In the Jaffe profile, it sits near the edge of the $10^2$–$10^4$ year time strip; with a Hernquist profile, it shifts to the $10^4$–$10^6$ year range; and under a Plummer profile, it moves even further out to the $10^6$–$10^8$ year strip. This outcome matches the intuitive expectation that a more extended and less centrally concentrated environment should respond more sluggishly to the evolving debris stream.

These comparisons highlight how strongly the structure of the host galaxy can shape the long-term evolution of TDE light curves. Very compact, cuspy systems like those described by the Jaffe profile tend to trigger environmental effects relatively early, but the resulting evolution remains smooth and rarely produces pronounced secondary peaks. Systems with intermediate concentration, such as Hernquist profiles, allow rebrightening to emerge over a wide portion of parameter space because they retain enough mass at intermediate radii to perturb the debris orbits more substantially. In contrast, diffuse cored environments like the Plummer sphere push these signatures to much later times, reflecting the slower dynamical influence of a more extended mass distribution. 

Looking at the lower panels, a pattern similar to what we discussed in Section \ref{sec:luminosity} shows up. Most of the parameter space is dominated by orange regions, indicating that the initial peak is super-Eddington, while any secondary features driven by the environment remain sub-Eddington. This behavior holds across all the profiles considered here. This means that for realistic systems, whether cored or cuspy, the environmental signatures almost never rival the initial flare in terms of luminosity. Only in extremely compact regions of the parameter space do these secondary features begin to approach or exceed the Eddington limit, but such regimes remain largely unobserved at present.


\section{Summary and Conclusions} \label{sec:conc}

With the growing number of observed TDEs, thanks to advances in time-domain astronomy, we are now seeing a far richer variety of light curve behaviors than classical models had predicted. Deviations from the standard $t^{-5/3}$ fallback rate and the appearance of multiple peaks in some events suggest that the disruption process may be more complex than previously assumed.

In this work, we revisited a key simplification made in most analytical TDE models, which is the assumption that the star evolves in a purely Keplerian potential dominated by the central black hole. While this may be valid for many galaxies, especially those with diffuse or extended stellar halos, we asked whether this still holds in more compact environments and in the later stages of debris evolution. To explore this, we developed a generalized framework that accounts for the gravitational influence of the host environment. Using the potential of a broken power-law density profile, we incorporated this external mass into the fallback dynamics and examined how it modifies the predicted light curve of the event.

Our model preserves the key features of the classical models, particularly the way stellar debris is spread in energy, but allows that energy to be shaped by the full galactic potential. This modification, though simple in its implementation, leads to significant changes in the resulting light curves, like the emergence of late-time features such as rebrightening episodes and periods of shallow fallback.

Through a detailed parameter study involving variations in the black hole mass, galactic mass, scale length and penetration factor, we arrive at several key findings:

\begin{enumerate}
    \item[(a)] Environmental effects tend to compete more effectively in systems with lower-mass black holes. In these cases, the surrounding galactic potential can more readily influence the fallback evolution, leading to earlier and more noticeable deviations, whereas in supermassive black hole systems, these signatures are typically suppressed.
    
    \item[(b)] The total galactic mass $m_{\mathrm{gal}}$ and scale radius $s$ strongly shape the light curve. Higher galactic mass amplifies deviations from the canonical fallback slope, while a larger scale radius delays their onset. Together, these parameters control both the timing and prominence of secondary features through their influence on the effective compactness of the environment.

    \item[(c)] The penetration factor also modulates the strength of the late-time environmental signatures, with deeper encounters producing them at lower fallback rates. The late-time slope evolution, however, remains largely insensitive to the encounter depth.

    \item[(d)] Environmental imprints generally appear at late times, well beyond the reach of current observations. Combinations of $(s, m_{\mathrm{gal}})$ that would produce early rebrightening (within $\sim 100$ years) require unrealistically compact systems in our framework.

    \item[(e)] The light curve is sensitive to the host galaxy’s density profile. Cuspy systems like Jaffe profiles tend to produce less pronounced secondary features. On the other hand, Hernquist and Plummer profiles, representing intermediate and more diffuse structures, are much more capable of generating delayed rebrightening.

    \item[(f)] Environmental signatures generally have sub-Eddington accretion rates, while the initial flare is typically super-Eddington. Only under extreme host compactness, can the environmental effect becomes comparable to or exceed the Eddington limit.
\end{enumerate}

Our results strongly suggest that the broader galactic structure can imprint itself directly onto TDE light curves. Many of these signatures, however, emerge only after the accretion rate has declined significantly, often at timescales far beyond what most current surveys can monitor. This could mean that some events thought to follow a simple monotonic decline are, in fact, incomplete light curves still evolving under the delayed influence of their host galaxy.

Environmental effects unfold over long timescales. This makes them difficult to pin down with current surveys, but it also opens an exciting opportunity. If we keep revisiting old TDE sites, some of them may still be evolving in ways we have not yet seen. Looking ahead, next-generation observatories with improved cadence, depth, and baseline will be better equipped to capture these elusive late-time features. When detected, they may not only provide new insights into the fallback process but also serve as indirect probes of galactic structure and black hole demographics.

It is also important to note that our treatment remains purely gravitational and Newtonian. We have not yet incorporated hydrodynamic processes within the debris stream or relativistic effects near the black hole, both of which can modify the overall light curve behavior. Including these ingredients will be necessary to fully assess how environmental effects compete or couple with fluid dynamics and strong-gravity corrections.

Several promising directions follow from this work. One immediate goal is to derive an analytic approximation which will complement the numerical results presented here. This will be presented in a separate paper. Beyond that, a more systematic scan of the parameter space, including those we held fixed here, may sharpen the predictive power of our current model. We have so far focused on spherically symmetric environments, but real galactic nuclei can be axisymmetric or even triaxial. Extending the framework to such configurations may expose new modes of debris evolution and behaviors absent in the spherical limit. Together, these efforts will help define the true reach of environmental effects by showing how strongly they can shape the late-time evolution of TDEs, how long their imprint can persist, and how much of the observed TDE diversity they can explain.


\appendix

\section{Potential of a double (broken) power-law density profile} \label{sec:appA}

The double (broken) power-law density distribution \citep{ah} is given by
    \begin{equation}
        \rho(r)=\rho_0\left(\frac{r}{b}\right)^{-\gamma}\left[1+\left(\frac{r}{b}\right)^\alpha\right]^{(\gamma-\beta) / \alpha}, \label{eq:A1}
    \end{equation}
with $\rho \propto r^{-\gamma}$ describing the density in the inner regions, while $\rho \propto r^{-\beta}$ at the outer regions. The $\alpha$ parameter determines the sharpness of the break. 

\vspace{5pt}

\noindent Here, we derive the potential associated with this density profile starting from the radially symmetric Poisson equation,
\begin{equation}
    \frac{1}{r^2} \frac{\partial}{\partial r}\left(r^2 \frac{\partial \Phi_{\mathrm{gal}}(r)}{\partial r}\right) = 4\pi G \rho(r). \label{eq:A2}
\end{equation}
The gravitational potential at radius $r$ due to a spherical mass distribution $\rho(r')$ is given by the sum of all the contributions from all the shells with $\mathrm{d} M\left(r^{\prime}\right) = 4 \pi \rho\left(r^{\prime}\right) r'^2 \mathrm{d} r^{\prime}$ inside and outside of $r$ as follows
\begin{equation}
    \Phi_{\mathrm{gal}}(r) = -4 \pi G(I_1+I_2), \label{eq:A3}
\end{equation}
where
\begin{subequations}
    \begin{align}
    I_1 &= \dfrac{1}{r} \int_0^r  \rho\left(r^{\prime}\right) r^{\prime 2}\mathrm{d} r^{\prime} \label{eq:A4} \\
    I_2 &= \int_r^{\infty}  \rho\left(r^{\prime}\right) r^{\prime}\mathrm{d} r^{\prime}. \label{eq:A5}
    \end{align}
\end{subequations}
We start by evaluating the integral $I_1$, we have
\begin{equation}
    I_1 = \frac{1}{r} \int_0^r \rho_0\left(\frac{r'}{b}\right)^{-\gamma}\left[1+\left(\frac{r'}{b}\right)^\alpha\right]^{(\gamma-\beta) / \alpha} r^{\prime 2} \mathrm{d} r^{\prime}. \label{eq:A6}
\end{equation}
Using the substitution $u=\left(\dfrac{r'}{b}\right)^{\alpha}$, we get
\begin{equation}
    \begin{aligned}
        I_1 &= \frac{b^3 \rho_0}{\alpha r} \int_{0}^{\left(\frac{r}{b}\right)^{\alpha}}  u^{\frac{3-\gamma-\alpha}{\alpha}}\Big(1+u\Big)^{\frac{\gamma-\beta}{\alpha}} \mathrm{d} u. \label{eq:A7}
    \end{aligned}
\end{equation}
Applying a second substitution $v=\dfrac{1}{1+u}$ then gives us
\begin{equation}
    \begin{aligned}
        I_1 &= -\frac{b^3 \rho_0}{\alpha r} \int_{1}^{\frac{1}{1+\left(\frac{r}{b}\right)^{\alpha}}} v^{\frac{\beta-3}{\alpha}-1}\left(1-v\right)^{\frac{3-\gamma}{\alpha}-1} \mathrm{d} v. \label{eq:A8}
    \end{aligned}
\end{equation}
Notice that the form of the above integral is similar to that of an incomplete Beta function
\begin{equation}
    \mathcal{B}(a, b, x)=\int_0^x t^{a-1}(1-t)^{b-1} \mathrm{d} t, \label{eq:A9}
\end{equation}
except that its lower bound is nonzero. We can resolve this issue by simply subtracting an incomplete beta function which runs from 0 to 1 (or simply the "complete" Beta function of the same integrand). We now have
\begin{equation}
    I_1 = -\frac{b^3 \rho_0}{\alpha r} \left\{\mathcal{B}\left(\frac{\beta-3}{\alpha},\frac{3-\gamma}{\alpha},\mathcal{X}\right) - \mathcal{B}\left(\frac{\beta-3}{\alpha},\frac{3-\gamma}{\alpha}\right)\right\}, \label{eq:A10}
\end{equation}
where
\begin{equation}
    \mathcal{X} = \frac{1}{1+\left(\frac{r}{b}\right)^{\alpha}}. \label{eq:A11}
\end{equation}
Similarly for $I_2$, we obtain
\begin{equation}
        I_2 = \frac{b^2 \rho_0}{\alpha} \mathcal{B}\left(\frac{\beta-2}{\alpha},\frac{2-\gamma}{\alpha},\mathcal{X}\right). \label{eq:A12}
\end{equation}
Combining these results for $I_1$ and $I_2$ gives us the following potential expression:
\begin{equation}
\begin{aligned}
    \Phi_{\mathrm{gal}}(r) = &\frac{4 \pi G \rho_0 b^3}{\alpha r} \left\{\mathcal{B}\left(\frac{\beta-3}{\alpha},\frac{3-\gamma}{\alpha},\mathcal{X}\right) \right.\\
    &\left.-\mathcal{B}\left(\frac{\beta-3}{\alpha},\frac{3-\gamma}{\alpha}\right) -\frac{r}{b}\mathcal{B}\left(\frac{\beta-2}{\alpha},\frac{2-\gamma}{\alpha},\mathcal{X}\right)\right\}. \label{eq:A13}
\end{aligned}
\end{equation}
Now, we simplify the above equation by solving for the total mass $M_{\mathrm{gal}}$ of the density distribution. Using the same substitutions as before, we should be able to get
\begin{equation}
    M_{\mathrm{gal}} = 4 \pi \int_{0}^{\infty} \rho\left(r^{\prime}\right) r'^2 \mathrm{d} r^{\prime} = \frac{4 \pi \rho_0 b^3}{\alpha}\mathcal{B}\left(\frac{\beta-3}{\alpha},\frac{3-\gamma}{\alpha}\right).
 \label{eq:A14}
\end{equation}
Expressing Equation \eqref{eq:A13} in terms of this total mass allows us to write the galactic potential more neatly as
\begin{equation}
    \begin{aligned}
        \Phi_{\mathrm{gal}}(r) =& -\frac{G M_{\mathrm{gal}}}{r}+ \frac{G M_{\mathrm{gal}}}{\mathcal{B}\left(\frac{\beta-3}{\alpha},\frac{3-\gamma}{\alpha}\right)} \left[\frac{1}{r}\mathcal{B}\left(\frac{\beta-3}{\alpha},\frac{3-\gamma}{\alpha},\mathcal{X}\right)\right.\\
    &\left.-\frac{1}{b}\mathcal{B}\left(\frac{\beta-2}{\alpha},\frac{2-\gamma}{\alpha},\mathcal{X}\right)\right]. \label{eq:A15}
    \end{aligned}
\end{equation}

\section{Newtonian (Keplerian) constraints on the maximum penetration factor}
\label{sec:appB}

Here, we estimate the physically allowed range of the
penetration factor for TDEs using Newtonian arguments. The penetration factor is defined as
\begin{equation}
    \delta \equiv \frac{R_{\mathrm{t}}}{R_{\mathrm{p}}},
\end{equation}
where $R_{\mathrm{t}}$ is the tidal radius and $R_{\mathrm{p}}$ is
the pericenter distance of the stellar orbit.

For a total tidal disruption to occur, the pericenter distance must
lie outside the capture radius of the black hole while remaining
within the tidal radius. This implies the allowed range
\begin{equation}
    1 \leq \delta \leq
    \frac{R_{\mathrm{t}}}{R_{\mathrm{cap}}},
\end{equation}
where the lower bound $\delta = 1$ corresponds to
$R_{\mathrm{p}} = R_{\mathrm{t}}$.

\subsection*{Capture radius and maximum penetration}

In Newtonian treatments, the smallest allowable pericenter is often
approximated either by the event horizon radius,
\begin{equation}
    R_{\mathrm{S}} =
    \frac{2 G M_{\mathrm{bh}}}{c^2},
\end{equation}
or more conservatively by the innermost stable circular orbit (ISCO),
\begin{equation}
    R_{\mathrm{ISCO}} = 3 R_{\mathrm{S}}
\end{equation}
for a non-rotating Schwarzschild black hole. Encounters with
$R_{\mathrm{p}} < R_{\mathrm{ISCO}}$ are typically unstable and
result in prompt capture rather than observable disruption.

Using the Newtonian tidal radius
\begin{equation}
    R_{\mathrm{t}} \approx R_* \left(\frac{M_{\mathrm{bh}}}{M_*}\right)^{1/3},
\end{equation}
the maximum penetration factor becomes
\begin{equation}
    \delta_{\mathrm{max}} = \frac{R_{\mathrm{t}}}{R_{\mathrm{cap}}} = R_*\left(\frac{M_{\mathrm{bh}}}{M_*}\right)^{1/3} \frac{c^2}{2 G M_{\mathrm{bh}}}.
\end{equation}

\begin{table}
\centering
 \caption{Maximum penetration factors for a Sun-like star
 for different black hole masses. Values are shown assuming
 the event horizon and ISCO as the limiting capture radius. The corresponding maximum stellar compactness allowing tidal disruption for each black hole mass is also listed.}
 \label{tab:delta}
 \begin{tabular}{c c c c c }
 \hline
  BH mass & $\delta_{\odot,\mathrm{crit}}$ &  $\delta_{\odot,\mathrm{crit}}$ &  $\mathcal{C}_{\mathrm{max}}^{-1}$ \\
  $\left(M_{\odot}\right)$ & (at event horizon) &  (at ISCO) &  \\
  \hline \hline
  $10^2$ & $10931.0$ & $3643.65$ & $21.5443$\\
  \\
  $10^3$ & $2355.00$ & $785.001$ & $100.000$ \\
  \\
  $10^4$ & $507.370$ & $169.123$ & $464.159$ \\
  \\
  $10^5$ & $109.310$ & $36.4365$ & $2154.43$ \\
  \\
  $10^6$ & $23.5500$ & $7.85001$ & $10000.0$ \\
  \\
  $10^7$ & $5.07370$ & $1.69123$ & $46415.9$ \\
  \\
  $10^8$ & $1.09310$ & No TDE & $215443$ \\
  \hline
 \end{tabular}
\end{table}

Introducing the stellar compactness parameter
\begin{equation}
    \mathcal{C} \equiv \frac{2 G M_*}{R_* c^2},
\end{equation}
this expression simplifies to
\begin{equation}
    \delta_{\mathrm{max}} = \mathcal{C}^{-1} \left(\frac{M_*}{M_{\mathrm{bh}}}\right)^{2/3}.
\end{equation}

This expression highlights the strong dependence of the maximum penetration factor on both stellar compactness and black hole mass.

\subsection*{Compactness limits}

A theoretical upper bound on stellar compactness is given by Buchdahl's limit,
\begin{equation}
    \mathcal{C} \leq \frac{8}{9}.
\end{equation}

Substituting this limiting value yields an absolute upper bound
\begin{equation}
    \delta_{\mathrm{crit}} = \frac{9}{8} \left(\frac{M_*}{M_{\mathrm{bh}}}\right)^{2/3}.
\end{equation}

However, stars approaching the Buchdahl limit are highly unrealistic for TDE scenarios. For astrophysical black hole masses, this bound generally produces $\delta_{\mathrm{crit}} < 1$, implying that the tidal radius would lie inside the Schwarzschild radius. In such cases, the star would be swallowed whole without disruption, preventing any observable TDE.

\subsection*{Application to Sun-like stars}

In our calculations, we consider a Sun-like star with compactness
\begin{equation}
    \mathcal{C}_{\odot} \approx 4.25 \times 10^{-6}.
\end{equation}

The maximum penetration factor for a solar-type star becomes
\begin{equation}
    \delta_{\odot,\mathrm{max}} = \mathcal{C}_{\odot}^{-1} \left(\frac{M_{\odot}}{M_{\mathrm{bh}}}\right)^{2/3}.
\end{equation}

Table \ref{tab:delta} lists the resulting maximum penetration factors for representative black hole masses, computed using both the event horizon and the ISCO as the limiting capture radius. The table also lists the maximum stellar compactness permitted for tidal disruption, which provides a useful measure of how dense a star can be for it be disrupted by a black hole of a given mass.


\acknowledgments

C.O. acknowledges the support of the University of the Philippines Diliman Office of the Vice Chancellor for Research and Development through Project No. 262616 TND.

\end{document}